\documentclass[12pt,aps,epsfig,amsmath,amssymb]{revtex4}
\usepackage{graphicx}
\usepackage{epsfig}

\begin{document}

\title{Hydrodynamic Theory for Reverse Brazil Nut Segregation
and the Non-monotonic Ascension Dynamics}
\author{
Meheboob Alam$^{1}$\footnote{
$^1$Engineering Mechanics Unit, Jawaharlal Nehru Centre for
Advanced Scientific Research, Jakkur P.O., Bangalore 560064, India;
Email: meheboob@jncasr.ac.in;
Ref. Journal of Statistical Physics, vol. 124, p. 587-623 (2006)
},
L. Trujillo$^{2}$
\footnote{$^2$Centro de F\'{i}sica, Instituto Venezolano de Investigaciones
Cent\'{i}ficas, A. P. Caracas 1029-A, Venezuela}
and H. J. Herrmann$^{3}$
\footnote{$^3$Institut f\"ur Computer Physik, Pfaffenwaldring 27,
Universit\"at Stuttgart,
D-70569 Stuttgart, Germany}
}
\affiliation{ 
}

\date{\today}

\begin{abstract}
Based on the Boltzmann-Enskog kinetic theory,
we develop a hydrodynamic theory for the well known (reverse) Brazil nut 
segregation  in a vibro-fluidized granular mixture.
Under strong shaking conditions, the granular mixture
behaves in some ways like a fluid and the kinetic theory  constitutive models
are appropriate to close the continuum balance equations for mass, momentum
and granular energy. 
Using this analogy with standard fluid mechanics, we
have recently suggested a novel mechanism of segregation 
in granular mixtures based on a {\it competition
between buoyancy and geometric forces}: the Archimedean buoyancy force,
a pseudo-thermal buoyancy force due to the difference between the energies
of two granular species, and two geometric forces, one compressive
and the other-one tensile in nature, due to the size-difference.
For a mixture of perfectly hard-particles with elastic collisions,
the pseudo-thermal buoyancy force is zero but the intruder has to
overcome the net compressive geometric force to rise.
For this case, the geometric force competes with the standard Archimedean buoyancy force
to yield a threshold density-ratio, $R_{\rho 1}=\rho_l/\rho_s < 1$,
above which the {\it lighter intruder sinks},
thereby signalling the {\it onset} of the {\it reverse buoyancy} effect.
For a mixture of dissipative particles, on the other hand,
the non-zero pseudo-thermal buoyancy force
gives rise to another threshold density-ratio, $R_{\rho 2}$ ( $> R_{\rho 1}$),
above which the intruder rises again.
Focussing on the  {\it tracer} limit of intruders in a dense binary mixture,
we study the dynamics of an intruder in a vibrofluidized system,
with the effect of the base-plate excitation
being taken into account through a `mean-field' assumption.
We find that the rise-time of the intruder could vary {\it non-monotonically}
with the density-ratio. 
For a given size-ratio, there is a threshold density-ratio
for the intruder at which it takes the maximum time to rise, and 
above(/below) which it rises faster,
implying that {\it the heavier (and larger) the intruder, the faster it ascends}.
The peak on the rise-time curve decreases in height and shifts
to a lower density-ratio as we increase the pseudo-thermal buoyancy force.
The rise (/sink) time {\it diverges} near the threshold density-ratio
for reverse-segregation. Our theory offers a {\it unified}
description for the (reverse) Brazil-nut segregation
and the non-monotonic ascension dynamics of  Brazil-nuts.
\end{abstract}

\maketitle

\newpage

\section{{\label{Intro}}Introduction}

The phenomenon of segregation, in which a homogeneous mixture of
particles of different species becomes spatially non-uniform by
sorting themselves in terms of their size and/or mass,
is ubiquitous in numerous chemical and pharmaceutical industries,
dealing with the transport and handling
of bulk granular mixtures \cite{HHL98,SegOld}. 
For most industrial processes, it is required to
maintain a homogeneous mixture during processing, with the segregation
being the `unwanted' phenomenon. Despite its unwanted
consequences, however, segregation occurs spontaneously in
driven granular mixtures
\cite{RSPS87,KJN93,DU,CWHB96,SM98,EXPNM,SIMU,SIMUCONV,
SW01,HQL01,BEKR03}.

When a mixture of different particles
is subjected to vertical shaking in a container,
the larger particles (intruders) rise to the free surface.
This is known as the {\it Brazil-nut phenomenon} (BNP),
and is one of the most puzzling phenomena of granular materials research,
still lacking a proper theoretical explanation.
The early experimental investigations on the Brazil nut phenomenon
\cite{SegOld,KJN93,DU,CWHB96} were
mainly concentrated on the effect of size
of the intruder particle in vibrated systems. Besides
experiments, computer simulations have been extensively used 
to study segregation \cite{RSPS87,SIMU,SIMUCONV,SW01}.
In fact, the current interest within the physics community on
granular segregation was stimulated by the Monte Carlo simulations
of Rosato and co--workers \cite{RSPS87}.
They explained BNP as a {\it percolation} effect 
since the smaller particles can easily percolate down to fill
the void, created behind the larger particles due to external shaking,
which, in turn, pushes the larger particle to the top.
However, we need to point out that such percolation effects 
are likely to be active in a dense bed only under {weak-shaking} conditions.

In many experimental setups convection is unavoidable
\cite{KJN93,CWHB96} and seems to be a key ingredient to
drive segregation \cite{KJN93,CWHB96,SIMUCONV}. 
In the regime of convection-driven segregation \cite{KJN93},
it has been experimentally verified that the large particles rise with
an {\it upward-plume} of the surrounding bed at the
center of the container, but cannot sink to the
bottom since they are unable to fit themselves in a
narrow {\it downward-plume} near the side-walls.
It has subsequently been verified, via careful MD simulations \cite{SIMUCONV},
that convection-driven segregation dominates (over percolation) in deep beds.
In this scenario
the intruder particles with different sizes rise at approximately the
same rate \cite{KJN93}.

However, a regime where convection
is negligible has  also been  found
in some experiments \cite{DU} as well as in many idealized
simulations \cite{RSPS87,SIMU,SW01}. The most striking observation
in this case is the dependence of the segregation-rate 
on the particle-size, where the ascending
velocity of the intruder particle increases with its size
\cite{DU,SIMU}. It has been  found that there is a {\it threshold size-ratio}
for the intruder above (/below) which the intruders rise (/sink).

A series of recent discoveries have shown that the Brazil nut phenomenon
is more intricate than it might seem. The subject took a major
step forward with the papers of Shinbrot and Muzzio \cite{SM98}
and Hong, Quinn and Luding \cite{HQL01}. In 1998 Shinbrot and
Muzzio \cite{SM98} discovered a {\it reverse buoyancy} effect: 
while a large heavy intruder rises to the free surface, 
an equally large but light intruder sinks to the bottom of
the granular bed. Three years later Hong, Quinn and Luding
\cite{HQL01} introduced a phenomenological theory for the {\it reverse Brazil nut
phenomenon} (RBNP): a competition between percolation
and condensation, depending on the size and the mobility of
the intruder particles, could drive the intruders 
to sink to the bottom of the container and vice versa.
The key idea of this theory is that a mono-disperse system of
particles can be fluidized above a critical temperature ($T_{cr}$),
assuming that the system is shaken uniformly.
Hence, if a binary mixture is vibrated  such that the mixture
temperature lies  between the critical temperatures of two individual species,
then one species will be in a fluidized state and the other condenses at the bottom,
leading to segregation.
Immediately the subject attracted the
attention of researchers who performed new experiments \cite{EXPNM,BEKR03}.

As mentioned before, different
mechanisms have been proposed to explain the segregation
phenomenon, for example, percolation \cite{RSPS87}, arching
\cite{DU}, convection \cite{KJN93,CWHB96,SIMUCONV}, inertia
\cite{SM98}, condensation \cite{HQL01}
and interstitial-fluid effects \cite{EXPNM}.
Unfortunately, while the observational evidence accumulates
\cite{SegOld,KJN93,DU,CWHB96,SM98,EXPNM,RSPS87,SIMU,SIMUCONV,SW01,BEKR03},
relatively little work exists on a unified theory for the dynamics of
Brazil nuts \cite{HQL01,DIFF,BH02,JY02}.
Recently, Jenkins and Yoon \cite{JY02} developed 
a theory for the segregation of elastic particles using
hydrodynamics equations of binary mixtures. 
They investigated the upward-to-downward transition introduced in
\cite{HQL01} even though their prediction of the
phase diagram for the BNP/RBNP-transition did not match with the simulation
results of Hong et al.
Interestingly, none of these early theories considered 
either the non-equipartition of granular energy
(which is a generic feature of granular mixtures)
or the effect of external driving
forces. Thus, it appears that a comprehensive theoretical description
for the dynamics of Brazil nuts is still lacking.

A hydrodynamic model to include the effect
of the non-equipartition of granular energy was postulated in ref. \cite{TH03} 
where a new mechanism for segregation due to buoyant forces
was introduced, drawing a direct analogy with the buoyancy forces in fluids.
More recently,  a {\it minimal} hydrodynamic model for segregation 
was outlined in ref. \cite{TAH03},
starting from the Boltzmann-Enskog-level
continuum equations for a dense binary mixture of fluidized
granular materials. The important effect of dissipation,
which is responsible for the
non-equipartition of the granular energy, was also incorporated. 
This latter work clearly shows how one could derive
a time-evolution equation for the relative rise velocity
of the intruder from first principles, 
thereby making the analogy with standard
fluids transparent (which has a history of more than a century, 
with the seminal works of Stokes, Oseen, Boussinesq, etc).
It was argued in ref. \cite{TAH03} that the BNP/RBNP segregation dynamics  results from
a competition between buoyancy and geometric forces.
This analysis \cite{TAH03} appears to be compatible with
the experimental observations reported by Breu {\it et al.}
\cite{BEKR03} and also agrees with the molecular dynamics simulations
reported by Hong {\it et al.} \cite{HQL01}.

In view of the good qualitative agreement of our model
\cite{TAH03} with the experimental observations presented by Breu
{\it et al.} \cite{BEKR03}, it is worth attempting to construct a
theory for segregation in granular materials using
the kinetic theory of inelastic hard particles.
We show that a {\it granular hydrodynamic} theory does indeed provide 
a good description for the dynamics of Brazil nut segregation.

This paper is organized as follows.
In Section \ref{KT} we introduce the kinetic theory for inelastic
hard particles, by summarizing the principal definitions of the
model and  presenting the balance equations for the mixture.
The problem of the non-equipartition of kinetic energy is considered in \ref{SKE}.
The main results of this paper are presented in Section \ref{ST},
where the detailed theory for segregation is outlined.
The evolution equation that governs the segregation dynamics
of intruders in the dense collisional regime is outlined in \ref{SD}.
We then comment on the {\it origin}
of different `gravitational' segregation
forces derived from the model in Section \ref{SF},
followed by a discussion on the {\it origin} of the unsteady forces that act on
the intruders in Section \ref{USF}.
In Section \ref{PD} we discuss the phase-diagram for the BNP/RBNP
transition in three-dimensions (i.e. for spherical particles),
thereby verifying that the 
competition between the buoyancy and geometric forces drives
the segregation process in a fluidized mixture.
In addition, we explain the {\it reverse buoyancy effect} of Shinbrot and 
Muzzio \cite{SM98} in Section \ref{RB}.
In Section \ref{DSI} we apply our theory to 
probe the dynamics of intruders in the tracer limit,
thereby explaining some recent experimental observations
on the {\it non-monotonic} rise-time (M\"obius et al. \cite{EXPNM}) of the intruder
(with the density-ratio) as well as 
the {\it divergence} of the rise(/sink) time
near the BNP/RBNP transition.
In Section \ref{sRBNP} we provide a simple analytical explanation for the
RBN effect, focusing on the Boltzmann (dilute) limit.
In Section \ref{HOE} we discuss the possible higher-order effects of 
the non-Maxwellian velocity distribution on segregation forces,
along with suggestions on some experimental implications of the present work.
The  conclusions and the limitations of our theory are detailed in Section \ref{CON}.

\section{{\label{KT}} Kinetic theory of granular mixtures}

We assume that granular matter 
can be described at the `macroscopic' level by a set of continuous
hydrodynamic equations as a fluid--mechanical medium.
The balance equations and the constitutive relations can be
derived from a kinetic theory description. For granular media
the kinetic equations are modified to account for the inelastic nature
of the collisions between particles\cite{JS83}. These equations have been
extended to binary mixtures \cite{JM87,WA99,GD99}.
The validity of the hydrodynamic approach even in the dense limit
has recently been justified via a comparison of the theory with
various experiments \cite{EXPKT}.

Theoretical and numerical studies for binary granular mixtures
have shown that the two components have different kinetic (fluctuation) energies
\cite{JM87,GD99,AL02a,AL03,BT02a}
which has also been  confirmed in
vibrofluidized experiments \cite{FM02}. Recently, Alam and
Luding \cite{AL02a,AL03} have demonstrated that a proper
constitutive model for granular mixtures must incorporate the
effect of the non-equipartition of granular energy. Therefore, in the
present model we consider this breakdown of equipartition on the
kinetic energy and this is a fundamental difference with previous
related studies \cite{HQL01,BH02,JY02}.

The following subsection  intends mainly to
establish some notational conventions used throughout the paper
and recall some basic definitions on granular hydrodynamics. Here
we follow the lines of refs. \cite{JM87,WA99} in the
introduction of the kinetic theory model and refs. \cite{AL02a,AL03} in
the discussion of the breakdown of the equipartition of the
kinetic energy.

\subsection{Definitions}

As a mechanical model for a granular fluid we consider a binary
mixture of slightly inelastic, smooth particles (disks/spheres)
with radii $r_i$ ($i=l,s$, where $l$ stands for large particles
and $s$ for small), mass $m_i$ in two or three dimensions ($d=2, 3$). 
The coefficient of restitution for collisions between particles is
denoted by $e_{ij}$, with $e_{ij}\leq 1$ and $e_{ij}=e_{ji}$.

The average macroscopic quantities are calculated by taking appropriate moments
of the corresponding microscopic (particle-level) 
property in terms of the single particle
velocity distribution function $f_i({\bf c},{\bf r}, t)$ for each
species. By definition $f_i({\bf c},{\bf r}, t)\mathrm{d}{\bf
c}\mathrm{d}{\bf r}$ is the total number of particles which, at time
$t$, have velocities in the interval $\mathrm{d}{\bf c}$ centered
at ${\bf c}$ and positions lying within a volume element
$\mathrm{d}{\bf r}$ centered at ${\bf r}$, i.e.,
\begin{displaymath}
   \int f_i({\bf c},{\bf r},t) \mathrm{d}{\bf c} \mathrm{d}{\bf r} = N_i .
\end{displaymath}
If the particles are uniformly distributed in space, so that $f_i$
is independent of ${\bf r}$, then the number density $n_i$
of species $i$ is
\begin{displaymath}
   n_i =\int  f_i({\bf c},{\bf r},t) \mathrm{d}{\bf c},
\end{displaymath}
and the total number density is $n = n_l + n_s$.
The species mass density $\varrho_i$ is defined by the product of
$n_i$ and $m_i$, and the total mixture density is
\begin{displaymath}
   \varrho= \varrho_l + \varrho_s = \rho_l\phi_l + \rho_s\phi_s,
\end{displaymath}
where $\rho_i$ is the material density of species $i$ and $\phi_i$
is the $d$--dimensional volume fraction for species $i$:
\begin{displaymath}
   \phi_i = \frac{\Omega_d}{d}n_i r_i^d,
\end{displaymath}
where $\Omega_d$ is the surface area of a $d$--dimensional unit
sphere.

The mean value of any quantity $\psi_i=\psi({\bf c})$ of a
particle species $i$, is
\begin{displaymath}
   \left< \psi_i({\bf c}) \right>\equiv \frac{1}{n_i} \int
         \psi_i({\bf c}) f_i({\bf c}) \mathrm{d}{\bf c}.
\end{displaymath}
The mean velocity of species $i$ is ${\bf u}_i=\left< {\bf c}_i
\right>$.
The mass average velocity  ${\bf u}$ of the
mixture is defined by
\begin{displaymath}
  {\bf u}\equiv\frac{1}{\varrho}(\varrho_l{\bf u}_l+ \varrho_s{\bf u}_s).
\end{displaymath}
The peculiar (fluctuation) velocity of species $i$ is
${\bf C}_i\equiv {\bf c}_i - {\bf u}$,
and the diffusion velocity is
${\bf v}_i\equiv \left< {\bf C}_i \right>={\bf u}_i - {\bf u}$.
The species``granular temperature'' is defined proportional to the
mean kinetic energy of species $i$
\begin{displaymath}
  T_i\equiv\frac{1}{d}m_i\left< {\bf C}_i\cdot {\bf C}_i \right> ,
\end{displaymath}
and the mixture temperature is
\begin{displaymath}
  T\equiv \frac{1}{n}(n_lT_l + n_sT_s) .
\end{displaymath}
Let us remark that this generalized notion of temperature is
introduced for a theoretical convenience to take advantage of a
thermodynamical analogy for granular materials,
and thereby postulating a higher-order field variable.
Even though the definition of thermodynamic variables for
non--equilibrium states is straightforward theoretically, the
thermodynamics of non--equilibrium states has always been a matter
of debate which we will not touch upon here.

\subsection{Mixture Granular Hydrodynamics}

The evolution of the granular system is governed by the well known
balance equations for the mixture  density, momentum and energy:
\begin{equation}
   \dot{\varrho} = - \varrho \nabla \cdot {\bf u},
\label{DENS}
\end{equation}
\begin{equation}
  \varrho \dot{{\bf u}} = -\nabla \cdot {{\bf P}} +
     \sum_{i=l,s}n_i {\bf F}_i,
\label{MOMEN}
\end{equation}
\begin{equation}
  \frac{d}{2}\dot{T} = T\nabla \cdot {\bf j} - \nabla \cdot {\bf q}
     - {{\bf P}} : \nabla {\bf u} + \sum_{i=l,s} {\bf j}_i \cdot
        {\bf F}_i -\mathcal{D},
\label{ENERG}
\end{equation}
where the overdot indicates  the convective derivative:
$\partial_t(\cdot) + {\bf u}\cdot \nabla(\cdot)$. 
Here, ${\bf P}$ is the mixture
stress tensor, ${\bf F}_i$ is the external force acting on the
particle, ${\bf q}$ is the mixture energy flux, ${\bf j}$ is the
diffusive mass-flux  and $\mathcal{D}$ is the total inelastic
dissipation rate.
These equations are rigorous consequences of the Enskog--Boltzmann
kinetic equation \cite{KTDS}, extended to inelastic particles
\cite{JM87,WA99}, and must be supplemented 
with respective constitutive equations for ${\bf {P}}$,
${\bf q}$, ${\bf j}$ and $\mathcal{D}$.
For  $e_{ij}=1$, the collisional dissipation rate vanishes
($\mathcal{D}=0$), and consequently  we recover the standard
energy balance equation for a mixture of elastic hard spheres \cite{KTDS}.

\subsubsection{Species Balance Equations and Fluxes}
                                                                                        
The species momentum balance equation is:
\begin{equation}
   \partial_t(n_i{\bf u}_i) + \nabla \cdot (n_i {\bf u}_i{\bf
    u}_i) = - \frac{1}{m_i}\nabla\cdot{\bf {P}}_i +
    \frac{n_i}{m_i}{\bf F}_i + {\bf \Gamma}_i,
\label{MBE}
\end{equation}
where ${\bf \Gamma}_i$ is the momentum source which arises 
due to the interaction between unlike particles and
$\sum_{i=l,s}{\bf \Gamma}_i = {\bf 0}$.
In the limit of small spatial inhomogeneities 
(i.e. first order in the gradients of the mean fields)
the species stress tensor ${{\bf P}}_i$ 
has, at the Navier-Stokes level, the standard Newtonian form
\begin{equation}
  {\bf P}_i = p_i {\bf I} + \mu_i({\bf\nabla u} + {\bf\nabla u}^T),
\end{equation}
where $p_i$ is the partial pressure of species $i$,
$\mu_i$ is the viscosity of of species $i$ and ${\bf I}$ the unit tensor.
The equation of state for the partial
pressure of species $i$ can be written as \cite{JM87,WA99}:
\begin{equation}
    p_i = n_i Z_i T_i, 
\label{eqn_state}
\end{equation}
where $Z_i$ is the ``compressibility'' factor of species $i$,
\begin{equation}
    Z_i \equiv 1 + \sum_{j=l,s}K_{ij},
\end{equation}
and
\begin{equation}
     K_{ij} \equiv \frac{1}{2}\phi_j \chi_{ij} (1+R_{ij})^d.
\end{equation}
Here $\chi_{ij}$ is the contact value of the radial distribution function 
and $R_{ij}=r_i/r_j$ the size ratio.
The radial distribution functions of granular systems
are often approximated by their
elastic counterpart (see, for example,
\cite{JM87,WA99} and for related issues \cite{PDF}).
In this paper we use the following  functions for disks \cite{MCSL71}:
\begin{equation}
   \chi_{ij} = \frac{1}{1-\phi} +
              \frac{9}{8}\frac{\phi_lR_{il}+\phi_sR_{is}}{(1+R_{ij})(1-\phi)^2},
\end{equation}
and for spheres:
\begin{equation}
   \chi_{ij} = \frac{1}{1-\phi} +
            \frac{\phi_lR_{il}+\phi_sR_{is}}{(1+R_{ij})(1-\phi)^2} \left[ 3 +
             \frac{2(\phi_lR_{il}+\phi_sR_{is})}{(1+R_{ij})(1-\phi)} \right].
\end{equation}

An explicit expression for the momentum source term can be written as
\cite{JM87}:
\begin{eqnarray}
   {\bf \Gamma}_i &=&
           n_iK_{ik}T\left[\left(\frac{m_k-m_i}{m_{ik}}\right)\nabla\left(\ln T\right)
          + \nabla\left[\ln\left(\frac{n_i}{n_k}\right)\right] \right. \nonumber\\
      & & \left. -\frac{4}{r_{ik}}\left(\frac{2m_im_k}{\pi m_{ik} T}\right)^{1/2}
            \left(\mathbf{u}_i-\mathbf{u}_k\right)\right],
\label{gammal}
\end{eqnarray}
where $m_{ik}=m_i + m_k$ and $r_{ik} = r_i + r_k$, with $i\neq k$.
In deriving the above expression, it has been assumed
that the single particle velocity distribution
function of species $i$ is a  Maxwellian at its own granular energy $T_i$:
\begin{equation}
   f_i ({\bf c}, {\bf z}, t) = n_i \left( \frac{m_i}{2\pi T_i}
            \right)^{d/2} \exp \left( - \frac{m_i({\bf c} - {\bf u}_i)^2}{2T_i}
              \right) .
\end{equation}
This represents the zeroth-order approximation for the distribution function which 
has recently been verified in molecular dynamics
simulations of vibrofluidized binary granular mixtures \cite{PCMP03}.
Higher-order corrections to the distribution function
would appear beyond the Euler-level description that we neglect 
in the present work (see discussion in VI.B).

The energy balance equation of species $i$ is
\begin{eqnarray}
    \partial_t(n_iT_i) + \nabla\cdot (n_i{\bf u}_i T_i) + \nabla
     \cdot {\bf Q}_i + {\bf {P}}_i:\nabla{\bf u} \nonumber \\
     - n_i{\bf F}_i \cdot {\bf v}_i - \frac{\varrho_i}{\varrho}{\bf
     v}_i \cdot (\nabla \cdot {\bf {P}}_i - n {\bf
     F})=&\mathcal{D}_i,
\label{ESBE}
\end{eqnarray}
and the mixture energy flux is defined as
\begin{equation}
   {\bf q} =  \sum_{i=l,s}n_i T_i + {\bf Q}_i,
\end{equation}
where ${\bf Q}_i$ is the energy flux of species $i$:
\begin{equation}
   { \bf Q}_i = \kappa_i \sqrt{T_i} \nabla T_i,
\end{equation}
and  $\kappa_i$ is the analog of thermal conductivity of species $i$.
The rate of kinetic energy dissipation of species $i$,
$\mathcal{D}_i$, is  \cite{BT02a}
\begin{equation}
\begin{split}
   \mathcal{D}_i & =
         \frac{\sqrt{2}d}{\sqrt{\pi}\Omega_d}\frac{m_i\phi_i}{r_i^{d+1}}
         \sum_{j=l,s}\chi_{ij}R_{ij}^{d}\left(1+R_{ji}\right)^{d-1}\phi_jM_{ji} \\
   &  \quad \times \left[ M_{ji}\left(1-e_{ij}^2\right)\left(\frac{T_i}{m_i}
             +\frac{T_j}{m_j}\right) 
        +  2\left(1+e_{ij}\right)\frac{T_i-T_j}{m_{ij}}\right]\left(\frac{T_i}{m_i}
           +\frac{T_j}{m_j}\right)^{1/2},
\end{split}
\label{dissII}
\end{equation}
where $M_{ji}\equiv m_j/m_{ij}$.
The total rate of kinetic energy dissipation is simply
$\mathcal{D}=  \mathcal{D}_l + \mathcal{D}_s$.
We can split eqn. (\ref{dissII})  into two terms:
\begin{itemize}
\item[1.] The inter--species collisional dissipation rate,
$\mathcal{D}_i^{I}$,
\begin{equation}
   \mathcal{D}_i^{I}  \equiv  \frac{\sqrt{2}d}{\sqrt{\pi}\Omega_d}
            \sum_j \chi_{ij}R_{ij}^{d}(1+R_{ji})^{d-1}M_{ji}^2 (1-e_{ij}^2) 
         \frac{\phi_i\phi_j}{r_i^{d+1}}
           \frac{T_i^{3/2}}{m_i^{1/2}}\left( 1 +\frac{m_iT_j}{m_jT_i}\right)^{3/2}.
\label{dissInter}
\end{equation}

\item[2.] The exchange collisional dissipation rate,
$\mathcal{D}_i^{E}$,
\begin{equation}
   \mathcal{D}_i^E  \equiv  \frac{2\sqrt{2}d}{\sqrt{\pi}\Omega_d}
        \chi_{ik} R_{ik}^d (1+R_{ki})^{d-1} M_{ik} M_{ki} (1+e_{ik}) 
         \frac{\phi_i
             \phi_k}{r_i^{d+1}}\frac{T_i^{3/2}}{m_i^{1/2}}
              \left(1-\frac{T_k}{T_i}\right)\left(
                  1+\frac{m_iT_k}{m_kT_i}\right)^{1/2},
\label{dissExc}
\end{equation}
with $k\neq i$.
\end{itemize}
Note that  $\sum_{i=l,s} \mathcal{D}_i^E = 0$,
and the {\it exchange}  term
$\mathcal{D}_i^E$ is a consequence of the non--equipartition assumption.
On the other hand, with the equipartition assumption 
($T_l=T_s=T$), $\mathcal{D}_i^E = 0$ and Eq. (\ref{dissInter}) reduces to:
\begin{displaymath}
    \mathcal{D}_i   =  \frac{\sqrt{2}d}{\sqrt{\pi}\Omega_d} \sum_j
     \chi_{ij}R_{ij}^{d}(1+R_{ji})^{d-1}M_{ji}^2 (1-e_{ij}^2)\\
    \frac{\phi_i\phi_j}{m_i^{1/2}r_i^{d+1}}
        \left(\frac{T}{M_{ji}} \right)^{3/2}.
\end{displaymath}

\subsubsection{{\label{SKE}}Non-equipartition of Granular Energy}

It is worthwhile now to comment on the breakdown
of the equipartition partition principle for a granular mixture. 
Physically the lack of energy equipartition is determined by the
different dissipation rates (\ref{dissInter}) and (\ref{dissExc}).
An obvious question is, ``How each species will attain a different
granular temperature?'' The answer to this question depends on the
balance between the external power injected into the system and the 
energy dissipation rates for the two species \cite{GD99,BT02a,AL02a,AL03}, for example,
in a vibrofluidized bed.

Before formulating a theory for segregation, it is important to
have an estimate of the granular energy ratio $T_l/T_s$
in a fluidized granular mixture.
An adequate description of the vibrofluidized granular system 
requires a careful analysis of 
the steady version of the species energy balance equations as described above.
To formulate the associated  boundary-value problem, 
however, we need to impose appropriate  boundary conditions
at the base-plate and the free surface of the  vibrofluidized mixture.
The problem of boundary conditions to be satisfied for
the granular hydrodynamic equations 
of a vibratory mono-disperse system has been studied \cite{BRMM00},
but such studies for  binary
and polydisperse mixtures do not exist.
Hence the complete solution of the boundary-value
problem for a vibrofluidized mixture is left to a future effort.

In the results presented below, we have rather used the
granular energy ratio of Barrat and Trizac \cite{BT02a}
which was derived for a randomly heated granular gas (see Section IV).
Their expression  has subsequently been verified in 
the molecular dynamics simulations \cite{PCMP03} on vibrated granular mixtures.
Paolotti {\it et al.} \cite{PCMP03} showed that 
the granular energy ratio  remains constant in the bulk of the mixture, 
with the variations of  $T_l/T_s$
with height being concentrated in two boundary layers near the base plate
and the free surface. The experiments \cite{FM02} 
also showed the existence of a constant $T_l/T_s$ in the bulk.

\section{{\label{ST}}Theory of Segregation}

To illustrate how the segregation dynamics can be described by a
hydrodynamic approach, we make  the following approximations:
\begin{itemize}
\item  
We assume that the granular medium is in a fluidized state.
The  fluidized state can be realized when the granular
material is vibrated strongly in the vertical direction, typically 
by a harmonic excitation 
$z_p(t)=A\sin (\omega t)$, with the amplitude $A$ and the
frequency $\omega=2\pi f$. The normalized acceleration parameter
($\Gamma \equiv A\omega^2/\mathrm{g}$) satisfies $\Gamma\gg 1$.

\item                                   
We assume that the system is in a regime where the bulk-convection 
and the air-drag can be neglected.

\item
We neglect the viscous stresses (as well as any stress-anisotropy) for the case where
there is no overall mean flow in the system. For example,
in a typical experimental realization of a vertically vibrated bed
the mean velocity is zero (${\bf u}= {\bf 0}$). 

\item                                                                                           
We impose horizontal periodic boundary conditions to make the
equations analytically tractable. Therefore, 
the hydrodynamic fields vary only along the vertical direction
($\partial/\partial z(\cdot) \neq 0$, but
$\partial/\partial x(\cdot)=0$ and  $\partial/\partial y(\cdot) = 0$).
     
\end{itemize}

\subsection{{\label{SD}}Segregation Dynamics: Evolution Equation}
                                                                                           
Here we  outline the derivation of the evolution equation for the
Brazil nuts.
With the above assumptions, the  equations for the evolution
of the granular mixture, (\ref{DENS})--(\ref{ENERG}), (\ref{MBE})
and (\ref{ESBE}), simplify considerably.
Our starting point here is the {\it inviscid} species  momentum balance
equations  (\ref{MBE}):
\begin{eqnarray}
    \partial_t u_l &=& -\frac{1}{\varrho_l}\partial_z p_l - \mathrm{g} 
       + \frac{1}{\varrho_l}\Gamma_l,  \label{WEL}\\
    \partial_t u_s &=& -\frac{1}{\varrho_s}\partial_z p_s - \mathrm{g} 
       + \frac{1}{\varrho_s}\Gamma_s.  \label{WES}
\end{eqnarray}
Subtracting Eq. (\ref{WES}) from Eq. (\ref{WEL}) we obtain 
\begin{equation}
   \varrho_l \partial_t u_l^r = - \partial_z p_l +
     \frac{\varrho_l}{\varrho_s}\partial_z p_s + \left(1 +
    \frac{\varrho_l}{\varrho_s} \right)\Gamma_l,
\label{eqn_evol1}
\end{equation}
where $u_l^r = u_l - u_s$ is the relative velocity of the larger
particles. Now we need expressions for $\partial_zp_l$ and
$\partial_z p_s$.

From the equation of state (\ref{eqn_state}),
we define the following weighted-ratio of two partial pressures \cite{JY02}:
\begin{equation*}
   \Theta(\phi_i, \phi_i/\phi_j, r_i/r_j, T_i/T_j) = \frac{p_s/n_s}{p_l/n_l} 
        = \frac{p_s}{p_l}\frac{n_l}{n_s},
\end{equation*}
which depends on the species volume fractions, the volume fraction ratio, the
size-ratio and the temperature ratio.
The partial derivative of $p_l$ is  calculated from
\begin{equation*}
\partial_z p_l=\frac{\partial}{\partial z}\left(\frac{p_s}{\Theta}\frac{n_l}{n_s}\right) 
         = \frac{p_s}{\Theta}\frac{\partial}{\partial z}\left(\frac{n_l}{n_s}\right)
            + \frac{n_l}{n_s}\left(\frac{1}{\Theta}\frac{\partial p_s}{\partial z} -
                    \frac{p_s}{\Theta^2}\frac{\partial \Theta}{\partial z}\right).
\end{equation*}
Substituting this in Eq. (\ref{eqn_evol1}) and using the
expression for $\partial_z p_s$ in Eq. (\ref{WES}) (and after
some algebraic manipulations), we arrive at the
following evolution equation for $u_l^r$:
\begin{eqnarray}
  \varrho_l \partial_t u_l^r &=& 
    n_l\left[m_s \left(
     \frac{Z_l}{Z_s}\frac{T_l}{T_s} \right) - m_l\right]\mathrm{g} +
     \left[ 1 + \frac{n_l}{n_s\Theta} \right]\Gamma_l \nonumber\\
   & & - p_l\partial_z\left[\ln\left(\frac{n_l}{n_s\Theta}\right)\right] 
    + n_l\left[m_s \left(
     \frac{Z_l}{Z_s}\frac{T_l}{T_s} \right) - m_l\right]\partial_t u_s.
\label{evo1}
\end{eqnarray}
In the following we focus on the tracer limit of this equation
for a mixture where the number density of intruder (larger) particles 
is much smaller that of the smaller particles, i.e. $n_l<<n_s$.
In this tracer limit ($n_l<<n_s$), 
the intruders are assumed to stay far away from each other (this, of course,
implies that larger and smaller particles are homogeneously
mixed initially) and they do not influence
each-other's motion; 
in fact, as explained by Lopez de Haro and Cohen \cite{SS82},
except for the mutual and thermal diffusion coefficients,
the transport coefficients of binary mixtures with one tracer component
can be obtained from the general expressions of corresponding transport coefficients
by taking the limit $n_l/n_s\to 0$.

For analytical simplicity,
we make two assumptions: (1) the global temperature of the bed $T$
does not vary with height (see, for example, the experiments in \cite{WHP01});
(2) the temperature ratio $T_l/T_s$ remains constant
in the bulk  which has also been verified in experiments \cite{FM02}
as well as  in simulations \cite{PCMP03}
of vibrated granular mixtures.
In the dense collisional regime of a mixture with $n_l<<n_s$ and
$\partial_z T=0$, the term associated with the momentum exchange in Eq. (\ref{evo1})
can be approximated by:
\[
  \left(1+\frac{n_l}{n_s\Theta}\right)\Gamma_l \approx \Gamma_l \approx
           p_l \left(\frac{T_s}{T_l}\right)\left[
           \partial_z\left(\ln\left(\frac{n_l}{n_s}\right)\right) 
       -\frac{4}{r_{ls}}\left(\frac{2m_lm_s}{\pi m_{ls} T}\right)^{1/2}
            \left({u}_l-{u}_s\right)\right].
\]
Furthermore, in this limit, it can be shown that
$Z_l/Z_s\approx\phi^2/2$ and hence
${\partial_z}\ln({Z_l T_l}/{Z_s T_s}) \approx {\partial_z}\ln\phi^2 = 2\lambda$,
where $\lambda$ is the decay rate of the volume fraction with height: 
$\ln\phi=\delta + \lambda z$. An approximate value for $\lambda$
can be estimated  from the vibrofluidized experiments \cite{WHP01}:
$\lambda r_s< -0.05$ (see Fig. 4$c$ in \cite{WHP01}) 
which is assumed to be valid in the dense regime.
Now, the third term on the right-hand side of  Eq. (\ref{evo1}) and the first term
in the above expression can be combined to yield
\begin{eqnarray*}
           p_l \left(\frac{T_s}{T_l}\right)
           \partial_z\left[\ln\left(\frac{n_l}{n_s}\right)\right] 
    - p_l\partial_z\left[\ln\left(\frac{n_l}{n_s\Theta}\right)\right]
&=& p_l\left(\frac{T_s}{T_l}-1\right)\partial_z\left[\ln
    \left(\frac{\phi_l}{\phi_s}\right)\right]
       + p_l\partial_z\left[\ln\left(\frac{Z_l T_l}{Z_s T_s}\right)\right] .
\end{eqnarray*}
Hence, both these terms are of order $O(\lambda p_l)$.
With these assumptions and retaining terms of same order in Eq. (\ref{evo1}), 
the time-evolution equation for the relative velocity of intruder particles 
$u_l^r$ is described by the following equation:
\begin{eqnarray}
 m_l\frac{{\rm d} u_l^r}{{\rm d} t} &=&
        \left[m_s\left(\frac{Z_l}{Z_s}\frac{T_l}{T_s}\right)- m_l\right] \mathrm{g}
        - \frac{4K_{ls}T}{r_{ls}}\left(\frac{2m_lm_s}{\pi m_{ls} T}\right)^{1/2}
     \!\!\!\!\!\! u_l^r \nonumber \\
   & &
   +\left[m_s\left(\frac{Z_l}{Z_s}\frac{T_l}{T_s}\right)-m_l\right]
           \frac{{\rm d} u_s}{{\rm d} t},
\label{dynamics}
\end{eqnarray}
where all expressions are evaluated in the limit $n_l/n_s\to 0$.

The above  equation (\ref{dynamics}) contains all the information necessary to describe
the segregation dynamics  of intruders in the tracer limit.
The first term on the right hand side is the net gravitational
force acting on the intruder.
Note that the second term has a form similar to the Stokes' drag
force, which always acts opposite to the intruder's movement.
The functional form of this term can be justified by recalling the experimental
observations of
Zik {\it et al.} \cite{ZSR92}, who found a linear dependence of
the drag force on the velocity of a sphere moving in a 
vibrofluidized granular medium.
This  could be interpreted as a characteristic of the
fluidized state of the granular media.
The last term represents a weighted {\it coupling} with the
inertia of the smaller particles. 
In the following subsection, we discuss the origin of 
these forces in detail.

\subsection{{\label{SegF}}Segregation Forces}

Let us now analyse different forces that are acting on the intruder
as it rises/sinks through the granular bed.

\subsubsection{{\label{SF}}Gravitational forces: buoyancy and geometric forces}
                                                                                           
We can decompose the net gravitational force in Eq.
(\ref{dynamics}) on an  intruder in the following
manner:
\begin{equation}
  F =  \textrm{g}
   \left[(\rho_s-\rho_l)V_l +
   m_s\left(\frac{T_l}{T_s}-1\right)\frac{Z_l}{Z_s}
   +  m_s\left(1-\frac{V_l}{V_s}\right)
       +m_s\left(\frac{Z_l}{Z_s}-1\right)\right],
\label{eqn_sforce1}
\end{equation}
where $V_i$ is the volume of a particle of species $i$.
This net gravitational force is composed of the following {\it
buoyant} and {\it geometrical} forces:
\begin{itemize}
                                                                                           
\item[$\bullet$] An {\it Archimedean buoyancy} force due to the
weight of
the displaced volume of the intruder  ($V_l$):
\begin{equation}
   F_b^A := V_l(\rho_s - \rho_l){\textrm{g}}.
\end{equation}
                                                                                           
\item[$\bullet$] An analogue of the {\it thermal buoyancy} force due to the
difference between the two granular temperatures:
\begin{equation}
  F_b^T := \beta (T_l - T_s),
\end{equation}
where $\beta=m_s T_s^{-1}(Z_l/Z_s)$ is the effective pseudo-thermal
expansion coefficient.
We call this {\it pseudo-thermal buoyancy} force.
Note that this force vanishes identically for a mixture of particles with elastic
collisions ($e_{ij}=1$).

\item[$\bullet$] 
Due to the size--disparity between the intruder and the smaller particles,
the intruder has to overcome a compressive volumetric strain,
$\epsilon_v^{st} :=(V_l/V_s-1)$.
This results in a {\it static compressive} force of the form:
\begin{equation}
  F_{ge}^{st} :=-m_s{\textrm{g}}\epsilon_v^{st},
\label{STen}
\end{equation}
{\it This force is always negative since the intruder has to rise against gravity}.

\item[$\bullet$] A {\it dynamic tensile} force from the pressure
difference due to the interaction between the intruder and the
smaller particles:
\begin{equation}
   F_{ge}^{dyn}:=m_s\textrm{g}\epsilon_v^{dyn},
\label{DTen}
\end{equation}
where
$\epsilon_v^{dyn}:=(Z_l/Z_s-1) \geq 0$
can be associated with a weighted {\it volumetric} strain, {\it
tensile} in nature.
\end{itemize}

The last two forces (\ref{STen}) and (\ref{DTen})  
are not related to standard buoyancy arguments.
Thus, purely {\it geometric} effects due to the {\it size-disparity}
contribute two new types of segregation forces:
\begin{equation}
  F_{ge} = F_{ge}^{st} + F_{ge}^{dyn} =
    - m_s\left(\epsilon_v^{st}-\epsilon_v^{dyn}\right)\textrm{g}.
\end{equation}
Overall, the collisional interactions
help to reduce the net compressive force that the intruder has to
overcome.
                                                                                           
It is interesting to find out whether we could get back the
standard Archimedes law from Eq. (\ref{eqn_sforce1}). 
This corresponds to the case
where a large particle is immersed in a sea of small particles with $r_l >> r_s$.
For this limit, it follows that \cite{TAH03} 
\begin{equation}
  F_{ge}^{dyn} \to m_s (V_l/V_s -1) = - F_{ge}^{st}
\end{equation}
and the net geometric force is 
\begin{equation}
   F_{ge} \equiv 0.
\end{equation}
Thus, the net gravitational force on a particle falling/rising in
an otherwise quiscent fluid (at the same temperature) 
boils down to
the Archimedean buoyancy force:
\begin{equation}
  F = F_B^A = \mathrm{g}(\rho_s - \rho_l)V_l.
\end{equation}

\subsubsection{{\label{USF}} Unsteady forces: the added mass effect}
                                                                                           
We have noted in the previous section that the inertia of the smaller particles
is directly coupled with the motion of the intruder 
in the evolution equation (viz. eq. \ref{dynamics}). 
For this unsteady (inertial) force $^tF$ also, we follow our earlier decomposition:
\begin{equation}
  ^tF = \frac{{\mathrm d}u_s}{{\mathrm d}t} 
      \left[(\rho_s-\rho_l)V_l +
        m_s\left(\frac{T_l}{T_s}-1\right)\frac{Z_l}{Z_s}
         + m_s\left(1-\frac{V_l}{V_s}\right)
         + m_s\left(\frac{Z_l}{Z_s}-1\right)\right].
\label{eqn_usforce1}
\end{equation}
It is evident now that the net unsteady term has contributions
from the standard {\it added mass force} along with two 
new forces as described below.
\begin{itemize}
\item
As in the unsteady-motion of a particle in a fluid,
the intruder in a granular mixture
has to rise along with its surrounding smaller particles. 
The `added' inertia
of the displaced smaller particles that are being carried
by the intruder gives rise to 
an effective  {\it added mass force} \cite{MR83} on the intruder:
\begin{equation}
   ^tF_{am} = V_l(\rho_s-\rho_l)\frac{{\mathrm d}u_s}{{\mathrm d}t}.
\end{equation}
As expected, this force vanishes if the material density
of the intruder is the same as that of smaller particles.

\item
A thermal analog to the {\it added mass force}, due to the
difference between the two  granular temperatures, is given by:
\begin{equation}
   ^tF_{am}^T = m_s\frac{Z_l}{Z_s}\left(\frac{T_l}{T_s}-1\right)
           \frac{{\mathrm d}u_s}{{\mathrm d}t}.
\end{equation}
This force vanishes identically for  $T_l=T_s$ that holds if the particle collisions
are perfectly {\it  elastic}. Hence this represents
a {\it new} force for the granular system.
 
\item
Lastly, we have  exact analogues of the two geometric forces as
described in the previous section:
\begin{equation}
   ^tF_{am}^{ge} = m_s\left(1-\frac{V_l}{V_s}\right)
           \frac{{\mathrm d}u_s}{{\mathrm d}t} 
     = -m_s\epsilon_v^{st}\frac{{\mathrm d}u_s}{{\mathrm d}t}.
\end{equation}
\begin{equation}
   ^tF_{am}^{ge} = m_s\left(\frac{Z_l}{Z_s}-1\right)
           \frac{{\mathrm d}u_s}{{\mathrm d}t}
            = m_s\epsilon_v^{dyn}\frac{{\mathrm d}u_s}{{\mathrm d}t}.
\end{equation}
While the former is the analog of static `compressive' geometric force,
the latter is the analog of the dynamic `tensile' geometric force
as discussed in the previous section.
Both these forces vanish if the intruder is of the same size
as the bed-materials.
\end{itemize}

Thus, we have shown that the last term in the evolution
equation (\ref{dynamics}) can be represented 
by an weighted {\it added-mass force}.

\section{\label{PD}Phase Diagram for BNP/RBNP transition and the Reverse Buoyancy Effect}

To proceed in the simplest possible way, we first  consider the {\it
steady-state} solution of  equation (\ref{dynamics})--
in this case the added-mass forces do not influence
the {\it onset} of segregation.
Neglecting transient effects, the
steady relative velocity of the intruder can be obtained from
\begin{equation}
   {u}_l^r = \frac{r_{ls}\textrm{g}}{4 K_{ls}}\left(\frac{\pi
             m_{ls}}{2m_lm_s T}\right)^{1/2}
             \left[m_s\left(\frac{Z_l}{Z_s}\frac{T_l}{T_s}\right)- m_l\right].
\label{Eq_SegCrit}
\end{equation}
Setting this  relative velocity to zero, we obtain the criterion
for the {\it transition} from  BNP to RBNP:
\begin{equation}
   m_s\left(\frac{Z_l}{Z_s}\frac{T_l}{T_s}\right) -m_l = 0,
\label{SegCrit}
\end{equation}
which agrees with the expression  of Jenkins and Yoon
\cite{JY02} for the case of equal granular energies ($T_l=T_s$),
i.e.   $Z_l/Z_s  = m_l/m_s$.
We have already noted in our previous paper \cite{TAH03} that 
the non-equipartition of granular energy ($T_l\neq T_s$)
that arises from the dissipative nature of particle collisions
must be incorporated into the theory 
to correctly describe many experimental findings.

As mentioned before, the energy ratio, $R_T=T_l/T_s$,
is calculated from the model of Barrat and Trizac \cite{BT02a}:
\begin{equation}
  C_1R_T^{3/2} + C_2\left(1+\frac{m_s}{m_l}R_T\right)^{3/2}
   +C_3\left(1+\frac{m_s}{m_l}R_T\right)^{1/2}\left(R_T-1\right) +C_4 =0,
\label{eqn_Gratio}
\end{equation}
where
\begin{eqnarray*}
 C_1 &=& 2^{d-1}(1-e_{ll}^2) \phi_l R_{sl}^d  \chi_{ll}\left(\frac{m_s}{m_l}\right)^{3/2}, \\
 C_2 &=& \sqrt{2}(1-e_{ls}^2)(1+R_{sl})^{d-1}\left(\phi_s M_{sl}^2, 
    - \phi_l R_{sl}^d M_{ls}^2\right)\chi_{ls}, \\
 C_3 &=& 2\sqrt{2}(1+e_{ls})(1+R_{sl})^{d-1}M_{sl}\left(\phi_s M_{sl} 
   + \phi_l R_{sl}^d M_{ls}\right)\chi_{ls}, \\
 C_4 &=& -2^{d-1} (1-e_{ss}^2) \phi_s R_{sl}^{d-1} \chi_{ss} .
\end{eqnarray*}
To draw the phase-diagram in the ($m_l/m_s, r_l/r_s$)-plane,
we need to solve to the segregation criterion
(\ref{SegCrit}) in conjunction with the expression for $T_l/T_s$ (\ref{eqn_Gratio}).
This leads to a quadratic polynomial for the mass-ratio $m_l/m_s$,
resulting in {\it multivaluedness} for $m_l/m_s$ below a
critical value of the size-ratio $r_l/r_s$ as we discuss in the following.
We note that another choice of $T_l/T_s$, as suggested by one of
the referees, does not change the qualitative nature of our results.

\begin{figure}
\includegraphics[width=10.0cm]{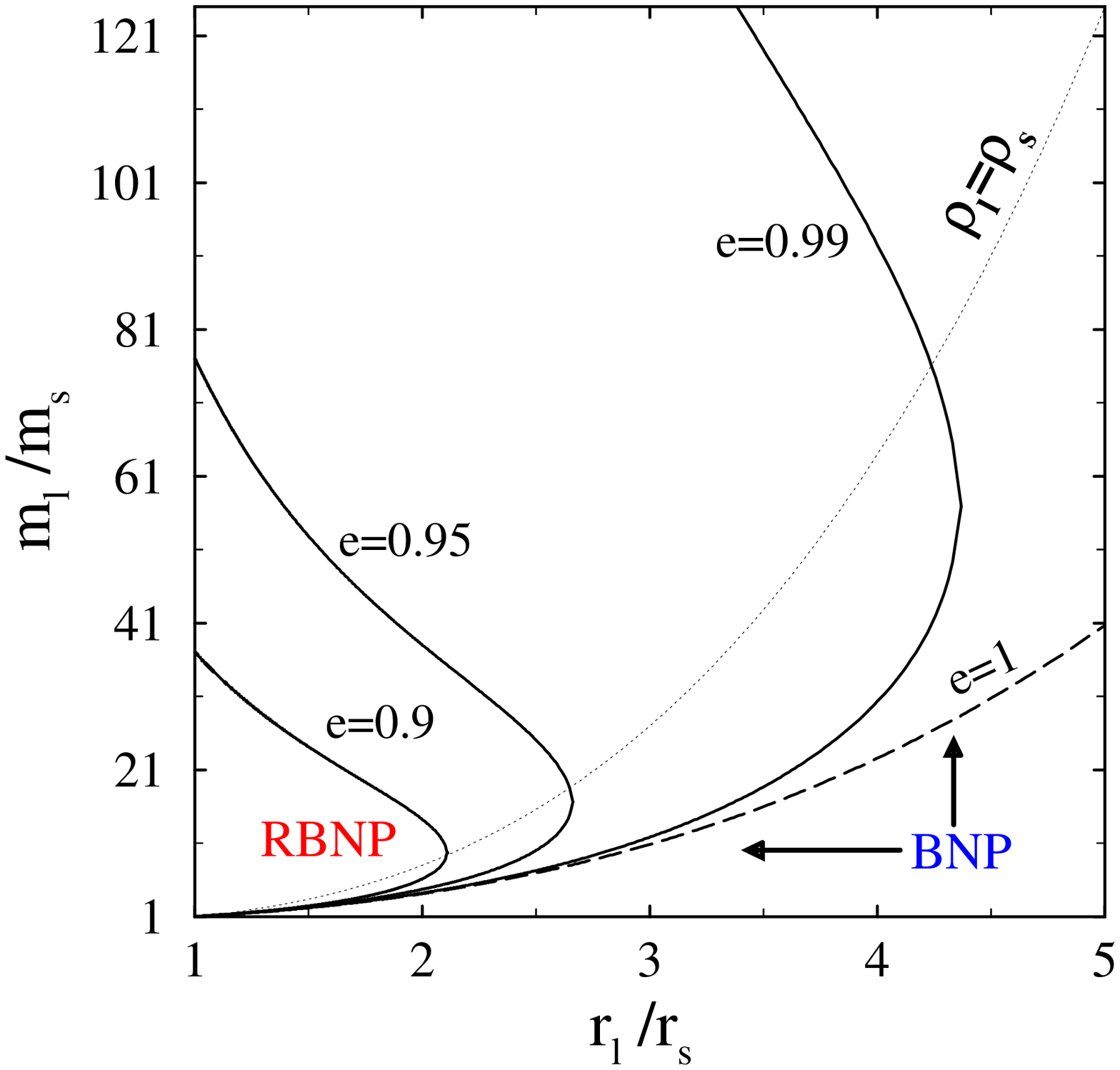}\\
\caption{
(color online)
Effect of inelasticity on the 
phase diagram for BNP/RBNP in three dimensions:
$\phi=0.5$ and $\phi_l/\phi_s=10^{-8}$.
The dotted line represents a mixture of equal
density particles ($\rho_l=\rho_s$): $m_l/m_s = (r_l/r_s)^3$.
}
\label{fig1}
\end{figure}

In Fig. \ref{fig1} we plot the phase diagram for the BNP/RBNP transition
for a mixture of spheres in the tracer  limit
($\phi_l/\phi_s=10^{-8}$) at a total volume fraction of $\phi=0.5$. 
Note that this volume fraction is well below the value
that corresponds to the perfect cubic-packing ($\phi=\pi/6\approx 0.52$),
implying that the mixture is in the `liquid' regime.
Each solid curve in Fig. \ref{fig1}, for a specified restitution coefficient,
demarcates the zones of BNP and RBNP transition.
We observe  that the qualitative nature 
of the phase-diagram changes even if 
the particles are {\it slightly} inelastic ($e=0.99$);
for example, the mass-ratio $(m_l/m_s)$ is a {\it multi-valued}
function of the size-ratio for $e\neq 1$, in contrast to the
perfectly elastic case (denoted by the dashed line). 
The effect of dissipation is to introduce a {\it threshold size-ratio}
above which there is no RBNP.
Moreover, the zone of  RBNP shrinks dramatically
when the particles are more dissipative; for this parameter
combination, there is no reverse segregation for  
moderately dissipative particles $e<0.8$.
We shall come back to discuss this point in the next
subsection in connection with the reverse-buoyancy
effect \cite{SM98}.

As discussed in our previous paper \cite{TAH03}, we need to
look at various segregation forces to understand the driving mechanism for
BNP/RBNP transition.
First, we focus our attention to a mixture of equal density particles 
($\rho_l=\rho_s$); this case is easily amenable to experiments by
using the intruder and the smaller particles of the same material.

\begin{figure}
\includegraphics[width=10.0cm]{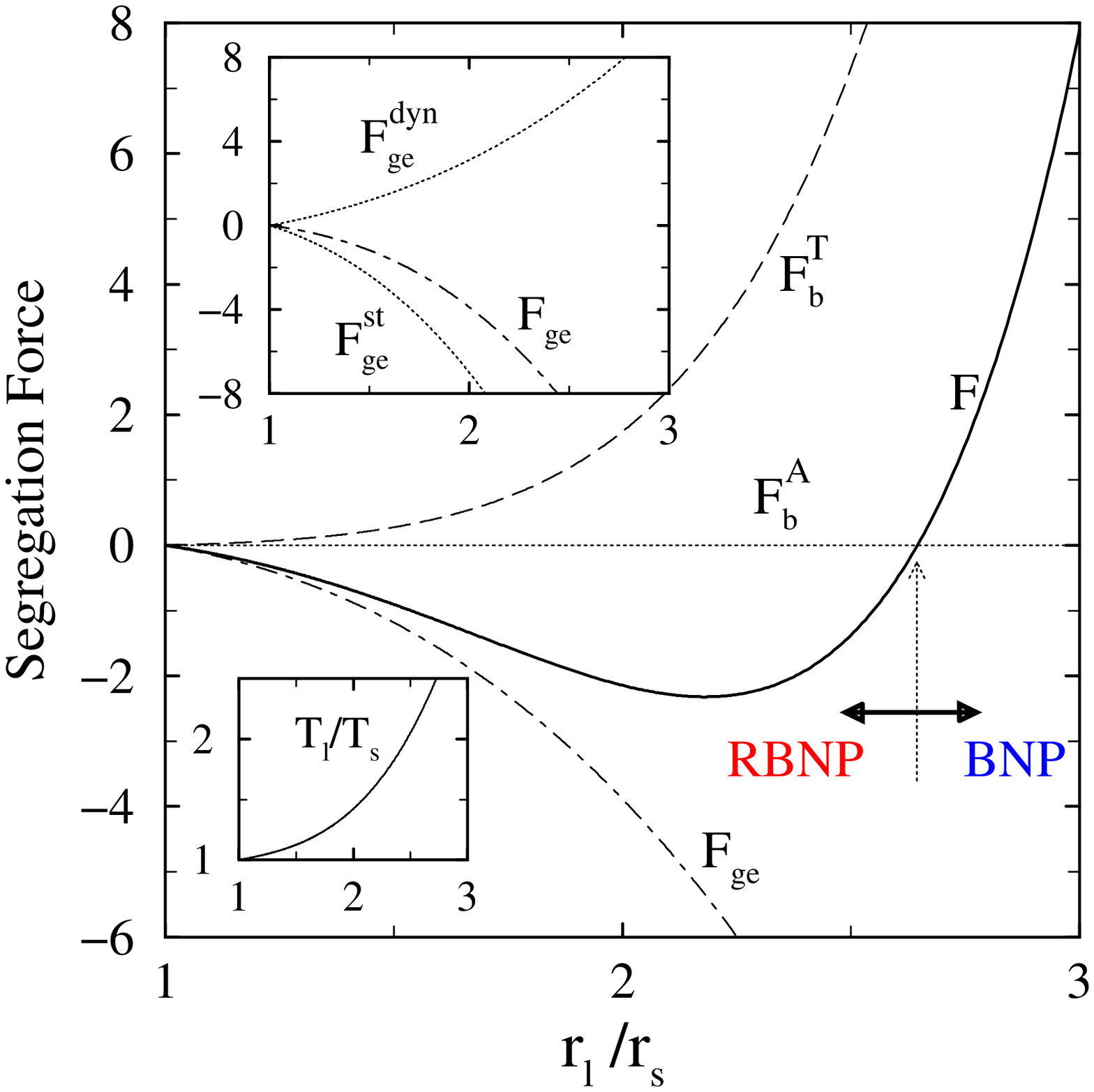}\\
\caption{
(color online)
Variations of segregation forces ($F/m_s g$) with the size-ratio for
a mixture of equal-density particles ($\rho_l=\rho_s$) at $e=0.95$.
The dotted arrow indicates the locus of the transition  
${\rm BNP \Leftrightarrow RBNP}$.
The upper inset shows the variations of the 
static and dynamic contributions to the total geometric force
($F_{ge}=F_{ge}^{st} + F_{ge}^{dyn}$) with the size-ratio.
The lower inset shows the variation of $T_l/T_s$ with the
size-ratio \cite{BT02a}.
}
\label{fig2}
\end{figure}

For an equal density mixture, the variations of different segregation forces
with the size-ratio are plotted in Fig. \ref{fig2}.
The coefficient of restitution is set to $0.95$.
Since $\rho_l=\rho_s$, the
Archimedean buoyancy force, $F_b^A \propto (\rho_l-\rho_s)$,
is identically zero as shown by the dotted horizontal line in Fig. \ref{fig2}.
The upper inset in Fig. \ref{fig2} shows the variations of two geometric
forces, and the net geometric force, $F_{ge}$, is {\it negative} as
shown by the dot-dash line.
However, the pseudo-thermal buoyancy force, $F_b^T \propto (T_l-T_s)$,
is positive and increases with increasing size-ratio.
Thus, the net gravitational force, $F=F_b + F_{ge}\equiv F_b^T + F_{ge}$,
can be positive/negative when $F_b^T$ greater/less than $F_{ge}$, 
respectively, with the equality
being the BNP/RBNP transition point. The solid line in Fig. \ref{fig2}
shows the variation of the net gravitational force $F$ that changes
sign at a size-ratio of $r_l/r_s\approx 2.6$ above which the intruder rises (BNP)
and below which it sinks (RBNP).
Clearly, this {\it threshold size-ratio} is decided by a competition
between the buoyancy forces and the geometric forces,
leading to the onset of BNP/RBNP transition \cite{TAH03}.
Note that for a mixture of particles with elastic collisions ($e=1$), 
the pseudo-thermal buoyancy force is $F_b^T=0$
and hence $F\equiv F_{ge} <0$; the intruder in such a mixture will, therefore, sink
for the parameter combinations of Figs. \ref{fig1} and \ref{fig2}.

\begin{figure}
\includegraphics[width=10.0cm]{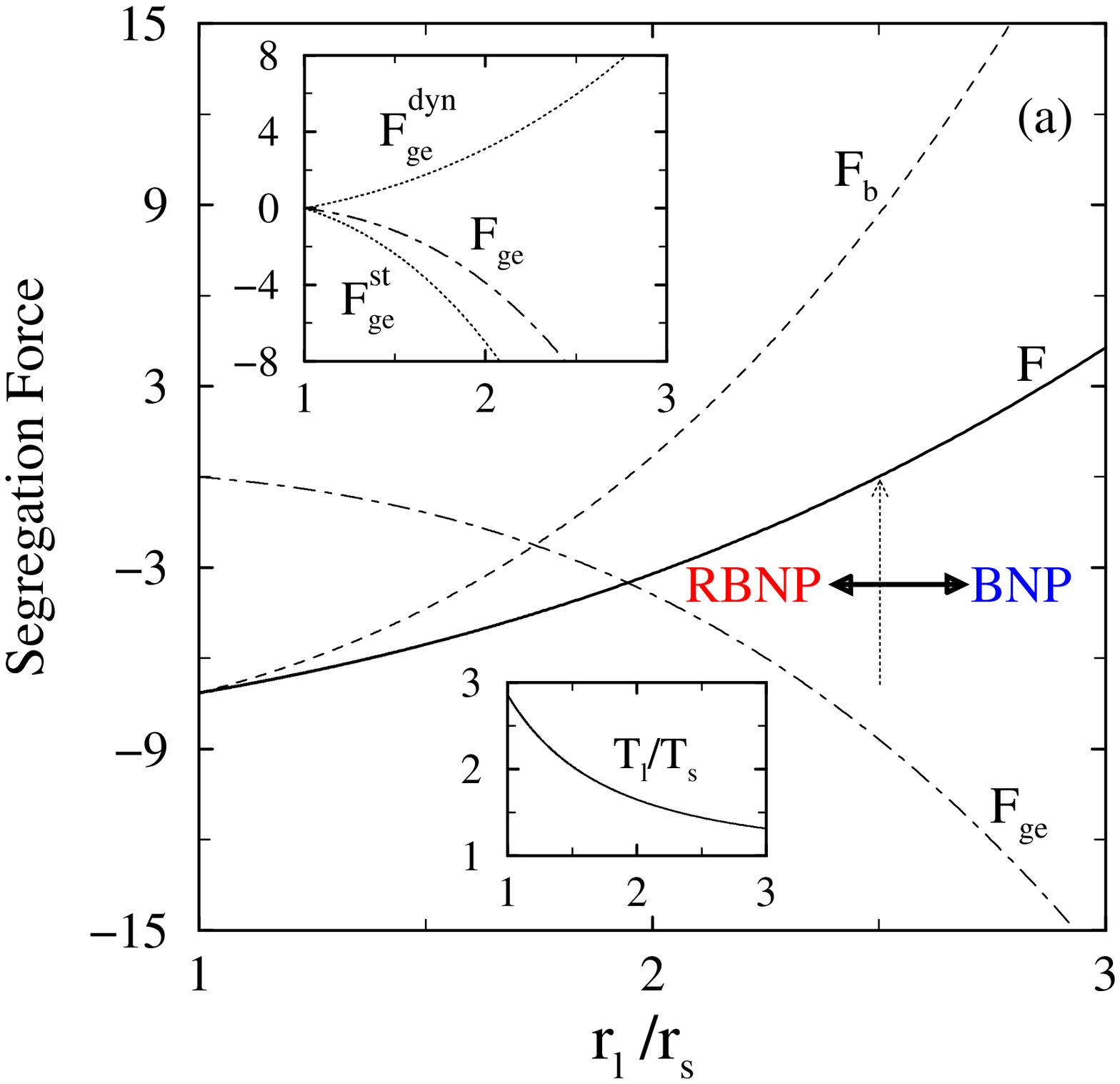}\\
\includegraphics[width=10.0cm]{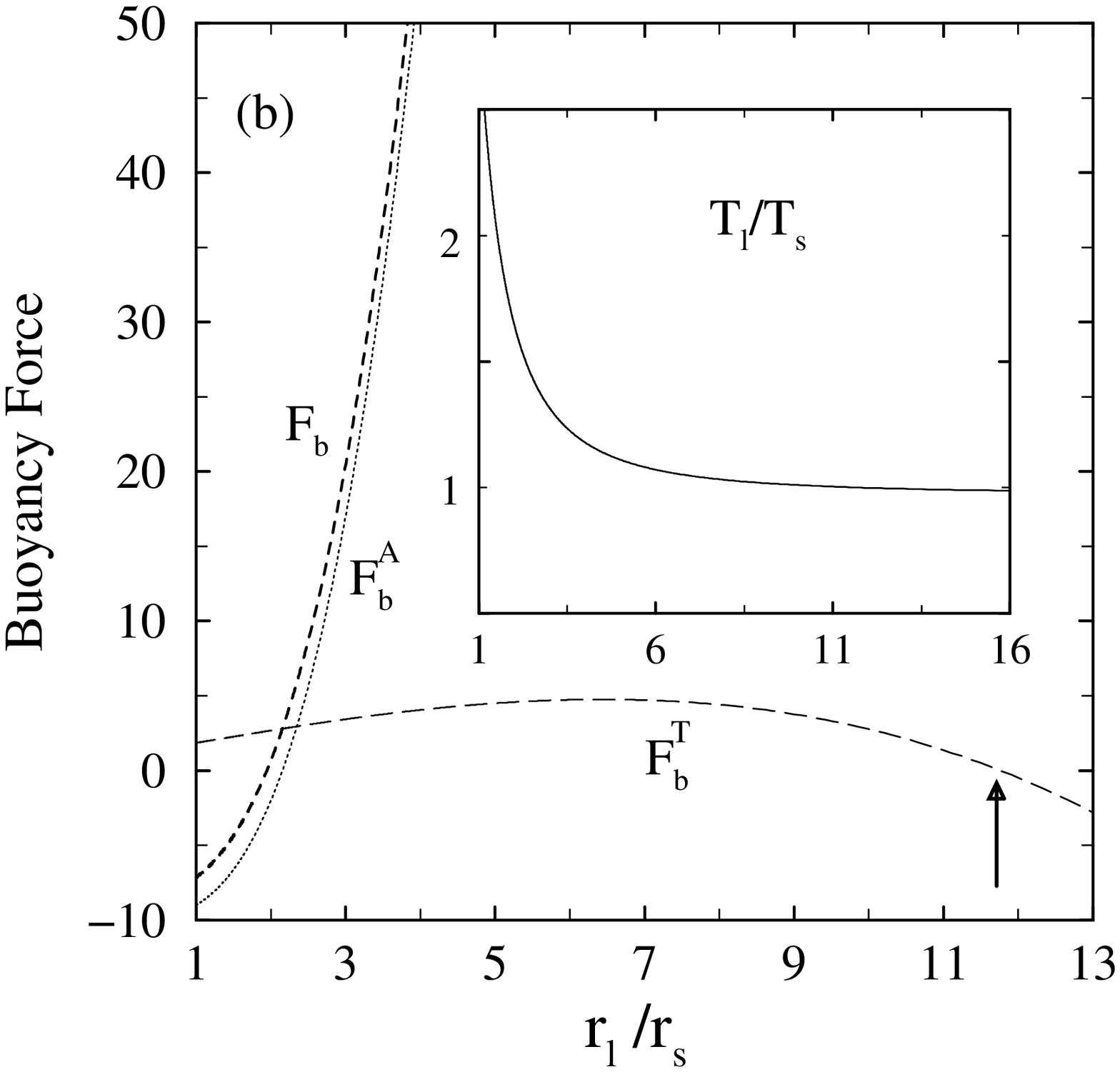}\\
\caption{
(color online)
(a) Variations of segregation forces ($F/m_s g$) with the size-ratio for
a mixture with fixed mass-ratio ($m_l/m_s=10$) at $e=0.95$.
The dotted arrow indicates the locus of the transition  
${\rm BNP \Leftrightarrow RBNP}$.
The upper inset shows the variations of the 
static and dynamic contributions to the total geometric force
($F_{ge}=F_{ge}^{st} + F_{ge}^{dyn}$) with the size-ratio.
The lower inset shows the variation of $T_l/T_s$ with the
size-ratio.
(b) Variations of Archimidean ($F_b^A$) and thermal ($F_b^T$)
buoyancy forces with $r_l/r_s$, with parameter values as in (a).
The arrow on the $F_b^T$-curve indicates the size-ratio
above which $F_b^T < 0$.
}
\label{fig3}
\end{figure}

Now we focus on a mixture with the mass-ratio, $m_l/m_s$,
between the intruder and the bed-particles being fixed;
this can be realized in experiments by fixing the bed materials
(glass beads or steel balls, etc.) and subsequently 
by varying the size and the density of the intruder.
The variation of segregation forces with the size-ratio
is shown in Fig. \ref{fig3} for $m_l/m_s=10$;
other parameters are as in Fig. \ref{fig1}.
Both the buoyancy and geometric forces are negative at $R_{ls}=1$
as seen in Fig. \ref{fig3}$a$. While the net buoyancy force increases
with the size-ratio, the net geometric force decreases in the same limit,
and again the competition between these two forces decides the onset of BNP/RBNP.
We observe in  Fig. \ref{fig3}$b$ that the pseudo-thermal buoyancy force $F_b^T$
remains positive only upto a size-ratio of $R_{ls}\leq 11.8$ and 
becomes negative thereafter (this is a consequence of the
energy-ratio $T_l/T_s$ being less than $1$ for $R_{ls} > 11.8$, 
see inset in Fig. \ref{fig3}$b$).
For this mixture, however, the Archimedean buoyancy force overwhelms its
pseudo-thermal counterpart (i.e. $F_b^A >> F_b^T$) beyond a moderate size-ratio 
of $R_{ls}>3$.
Hence, it is the competition between the Archimedean buoyancy force
and the geometric forces that mainly determines 
the {\it threshold size-ratio} for the BNP/RBNP transition ($R_{ls} \approx 2.5$) for 
this case.

\subsection{{\label{RB}}Reverse Buoyancy and Beyond}

In 1998 Shinbrot and Muzzio \cite{SM98} discovered 
an interesting effect: even though large heavy intruders can rise
to the top in a vibrofluidized mixture of smaller particles,
relatively {\it lighter} intruders of the same size can sink
to the bottom. This is in contradiction to common expectation
and has been appropriately dubbed the {\it reverse buoyancy effect}.
(Strictly speaking, however, the rising-phenomenon of
`heavier' intruders is also a reverse-buoyancy effect.) 
In the following we refer to the sinking-phenomenon of `lighter'
intruders ($\rho_l/\rho_s <1$) as the reverse-buoyancy effect 
as in Shinbrot and Muzzio.

\begin{figure}
\includegraphics[width=10.0cm]{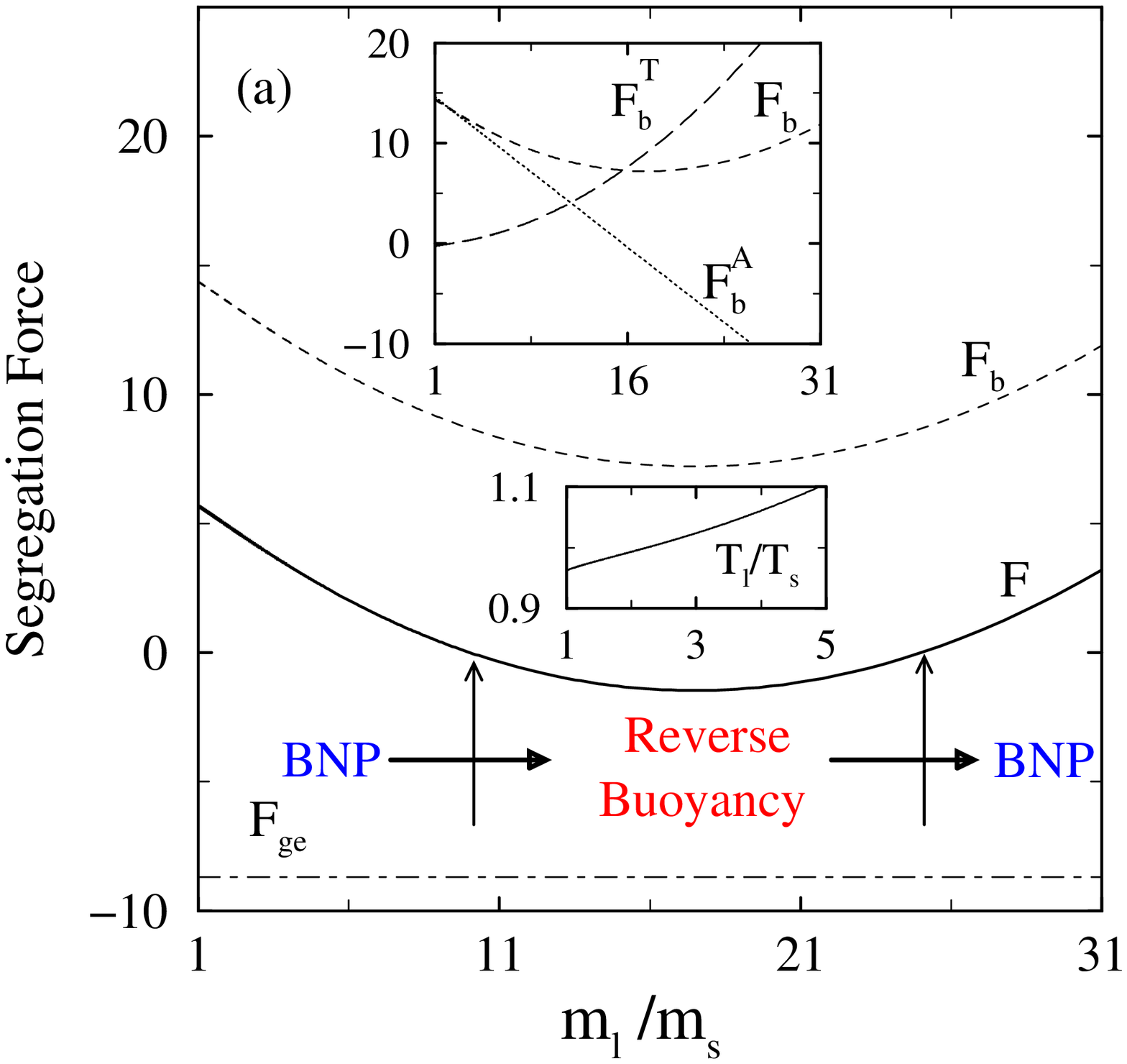}\\
\includegraphics[width=10.0cm]{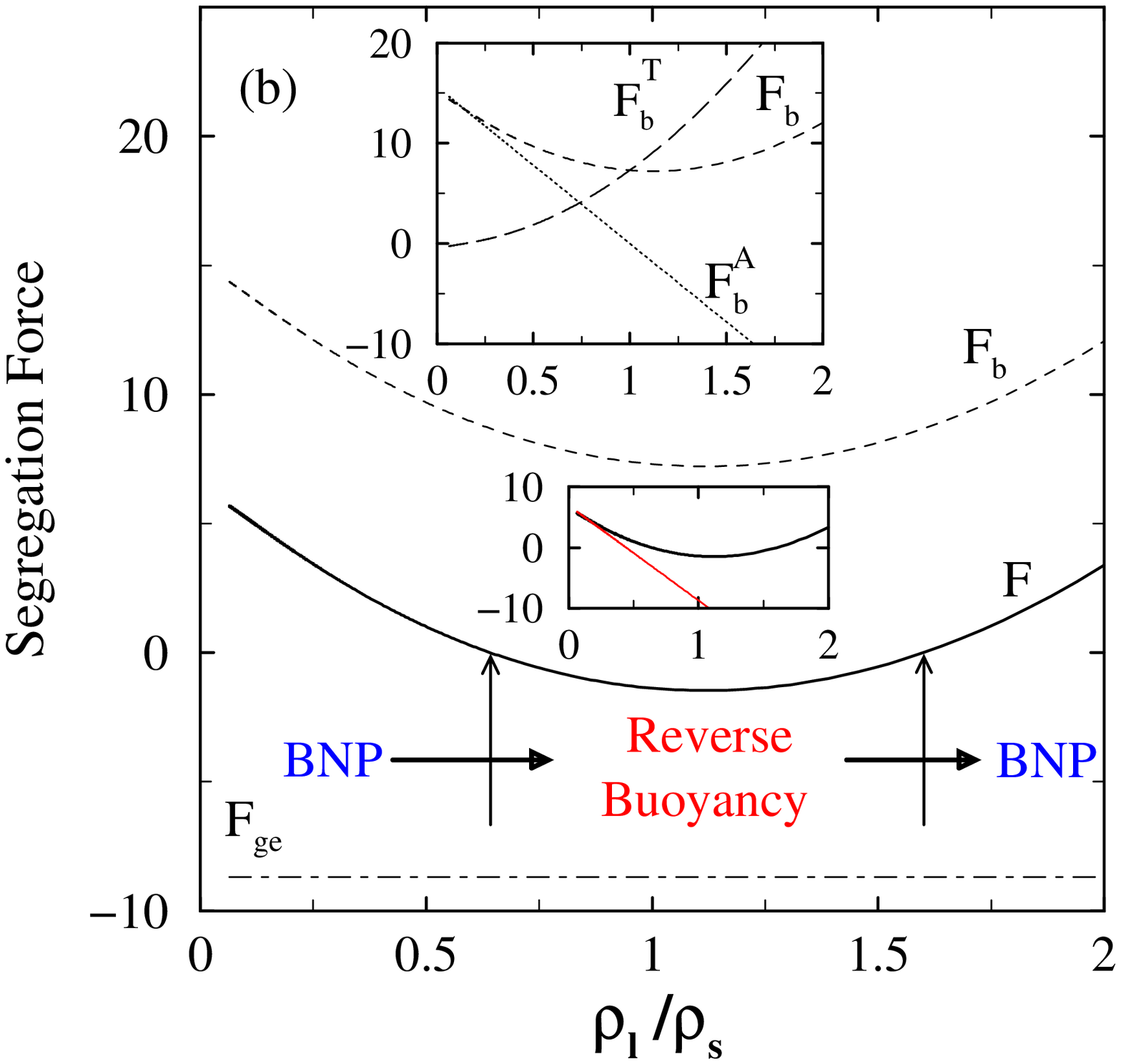}\\
\caption{
(color online)
(a) Onset of {\it reverse buoyancy effect} in terms of segregation forces:
variations of $F/m_s g$ with the mass-ratio for
a mixture with $r_l/r_s=2.5$.
Parameter values are $\phi=0.5$, $\phi_l/\phi_s=10^{-8}$ and   $e=0.95$.
The vertical arrows indicate the loci of the transition  
${\rm BNP \Leftrightarrow RBNP}$.
The upper inset shows the variations of the 
Archimedean and pseudo-thermal contributions to the total buoyancy force
($F_b=F_{b}^{A} + F_{b}^{T}$) with the mass-ratio.
The lower inset shows the variation of $T_l/T_s$ with the
mass-ratio.
(b) Same as in $a$ but with density-ratio, $\rho_l/\rho_s$.
The redline in the lower inset shows the variation of $F_b^A+F_{ge}$.
}
\label{fig4}
\end{figure}

To explore whether our model is able to explain
this effect or not, we have fixed the size-ratio at $R_{ls}=2.5$,
and changed  the mass-ratio by varying the density-ratio;
again this can be realized in experiments by varying the material
density of either the intruder or the smaller particles.
Figure \ref{fig4}$a$ shows the 
onset of the {\it reverse buoyancy effect} where
we have plotted the variations of different segregation forces
with the mass-ratio, $m_l/m_s$,
with other parameter values as in Fig. \ref{fig2}.
To contrast the mass-ratio and density-ratio effects simultaneously,
the same figure is redrawn in Fig. \ref{fig4}$b$, with
the abscissa now representing the density-ratio, $\rho_l/\rho_s$.
For a fixed size-ratio, the net geometric force remains constant
as shown by the dot-dash line in Fig. \ref{fig4}.
The upper inset in Fig. \ref{fig4} shows the 
Archimedean and pseudo-thermal contributions to the total buoyancy force;
while $F_b^T$ increases with the mass-ratio, $F_b^A$
decreases in the same limit, and the total buoyancy force
varies {\it non-monotonically} with both mass- and density-ratios.

Focussing on the mass-ratio of $m_l/m_s=30$ 
(the density-ratio is $\rho_l/\rho_s\approx 1.98$) in Fig. \ref{fig4},
we note that the net gravitational force is positive, i.e.
the heavier intruder will rise to the top (which is the BNP).
Looking at the upper inset in Fig. \ref{fig4}$a$, we find
that it is the pseudo-thermal buoyancy force that drives the 
intruder to the top (at such large density-ratios).
Now if we decrease the mass of the intruder (by decreasing its
material density $\rho_l$) to $m_l/m_s\approx 25$ (that
corresponds to a density-ratio of $ \rho_l/\rho_s\approx 1.6$),
the net gravitational force becomes negative.
Hence this relatively {\it lighter} intruder ($ \rho_l/\rho_s < 1.6$) 
will now sink to the bottom (i.e. which is the RBNP).
When we lower the density-ratio even below unity,
the lighter intruder sinks to the bottom as observed in Fig. \ref{fig4}$b$.
This is nothing but the reverse-buoyancy effect of Shinbrot and Muzzio \cite{SM98}:
{\it although the heavier intruder can rise to the top, equally large  
but a relatively {\it lighter} intruder can  sink}.

It is interesting to observe in Fig. \ref{fig4}
that there is a window of mass- and density-ratios 
($10 <m_l/m_s < 25$ and $0.64 < \rho_l/\rho_s < 1.6$)
for which the net gravitational force remains negative
and hence the intruder will sink.
For $\rho_l/\rho_s < 0.64$, however, the Archimedean buoyancy
force prevails over other forces and the lighter intruder rises to
the  top (as observed in the upper inset of Fig. \ref{fig4}$b$).
This implies that if the material density of the intruder
is much less than that of the bed particles, the intruder will
eventually show the standard buoyancy effect (which is again the BNP).

To understand the origin of reverse buoyancy, we 
consider  particles with perfectly
elastic collisions ($e=1$), for which $T_l=T_s$ and  $F_b^T=0$; hence
the net gravitational force, $F\equiv F_b^A + F_{ge}$, will decrease
monotonically with increasing size-ratio as shown by the red-line in the
lower inset of Fig. \ref{fig4}$b$. 
In such a mixture of elastic particles,
an intruder with  $R_{\rho 1}=\rho_l/\rho_s > 0.44$ will sink
(i.e. the reverse buoyancy),
and this {\it threshold density-ratio} is less than unity since
the lighter intruder has to overcome the net {\it compressive} geometric force.
Thus, {\it the onset of reverse buoyancy results from a competition
between the Archimedean buoyancy force and the net
compressive geometric force}.

The effect of the pseudo-thermal buoyancy force (for
dissipative particles) is simply to increase this  threshold density-ratio 
to $R_{\rho 1}\approx 0.64$ for parameter combinations of Fig. \ref{fig4},
and create another threshold density-ratio at $R_{\rho2}\approx 1.6$, above unity,
beyond which the heavier intruder rises to the top (i.e. BNP).
Thus, the lower threshold density-ratio $R_{\rho1}$ is 
created by the geometric forces and the higher threshold density-ratio $R_{\rho2}$
is created by the pseudo-thermal buoyancy force.

It is interesting to point out that the threshold density-ratio $R_{\rho2}$
at which the RBNP  occurs depends crucially on the
overall mean volume fraction (that can be related  with the shaking
strength of the vibrator, see Section \ref{ESS}) of the fluidized bed. 
For example, by reducing
the mean volume fraction, $R_{\rho2}$ can be pushed below unity.

Thus, our model predicts that the intruder will show
the BNP for $\rho_l/\rho_s >> 1$ and $\rho_l/\rho_s << 1$.
While the BNP in the former limit is driven by the
pseudo-thermal buoyancy force, the BNP in the latter limit
is driven by the Archimedean buoyancy force.
For intermideate values of density-ratios, the intruder will
show the RBNP (and the reverse buoyancy) which is driven by a competition between the 
buoyancy and geometric forces.

Now we  make a few comments on the experimental
realizations of reverse buoyancy.
The range of size-ratios for which 
our model predicts the reverse buoyancy effect
is of order $R_{ls}=O(5)$ for the coefficient of restitution $e<0.99$;
for larger values of $R_{ls}$
our model predictions show the standard Brazil-nut segregation.
However, this critical size-ratio can be pushed to a much higher value 
by considering higher values of $e$; this 
is equivalent to reducing the magnitude of the pseudo-thermal buoyancy ($F_b^T$),
since $F_b^T$ vanishes as the energy ratio $T_l/T_s$ approaches
its equipartition value for perfectly elastic system ($e=1$).

More recently, Yan {\it et al.} \cite{Yan03} 
have also observed the reverse buoyancy effect for 
a range of size-ratios varying between $25$ to $36$;
the diameter of their bed-particles was less than $250\mu m$.
Interestingly, the critical density at which this transition
occurred was  $R_{\rho2}=\rho_l/\rho_s\approx 0.7$.
But the interstitial
air-pressure has played a crucial role in their experiments
since {\it they could not observe RBNP when the
experiments were performed at a reduced  air-pressure} ($0.1$ atm).
Thus, the missing link could be provided by the
{\it air-drag} that we have neglected in our analysis.

Our predictions of a transition to BNP for very light intruders 
(at $R_{\rho}<< R_{\rho1}$)
has been recently observed  by Huerta and Ruiz-Suarez \cite{HR04}.
Their experiments at high frequencies 
($f=50$ Hz and $\Gamma=3$) would closely mimic most of our
assumptions since  bulk-convection was negligible in their experiments.
More careful experiments at vacuum are needed to 
map out the whole phase-diagram.

\subsection{{\label{ESS}}Effect of Shaking Strength}

Here we consider the effect of the shaking strength of  vibration 
on the BNP/RBNP transition.
In typical vibrofluidized-bed experiments, the mixture is vibrated via
a sinusoidal excitation of the base-plate:
\begin{equation}
   z_p(t) = A\sin(2\pi f t),
\end{equation}
where $A$ is the amplitude of vibration and $f$ is its frequency.
The shaking strength of vibration can be measured via
the following non-dimensional number
\begin{equation}
   \Gamma = \frac{A(2\pi f)^2}{g} .
\label{par_1}
\end{equation}
Ideally, the  fluidized regime corresponds to both large-amplitude
and high-frequency vibrations which can be related to the case $\Gamma>> 1$.
Note that this condition, $\Gamma>> 1$, can be
realized either by increasing the shaking amplitude at fixed $f$
or by increasing the shaking frequency at fixed $A$.

\begin{figure}
\includegraphics[width=10.0cm]{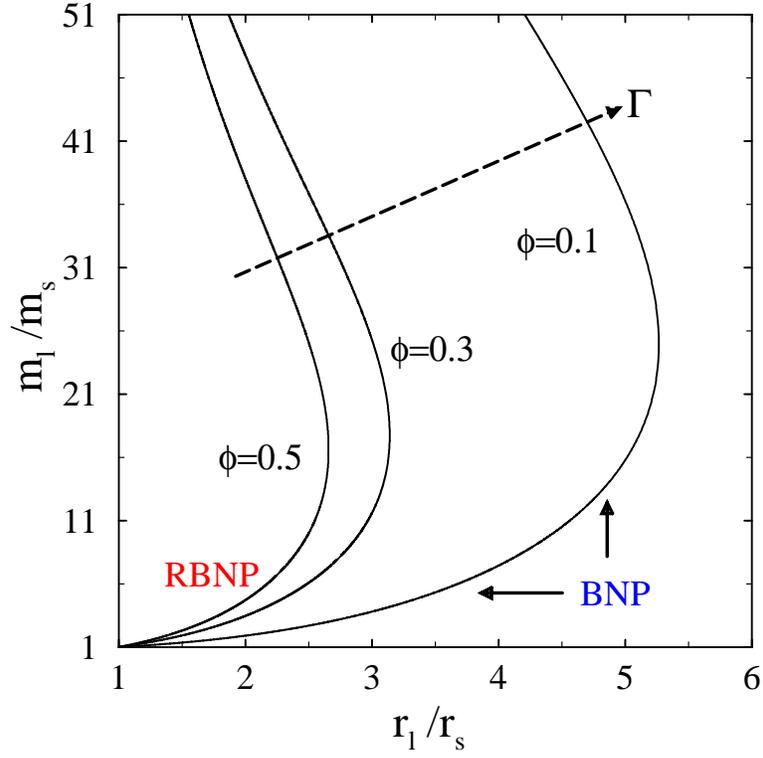}\\
\caption{
(color online)
Effect of mean volume fraction on the 
phase diagrams for BNP/RBNP in three dimensions:
$e=0.95$ and $\phi_l/\phi_s=10^{-8}$.
The dashed arrow indicates the direction of increasing shaking
strength $\Gamma$.
}
\label{fig5}
\end{figure}

In the fluidized-regime, the bed expands with increasing shaking strength $\Gamma$,
leading to a decrease in the overall volume fraction of the mixture $\phi$.
Thus, the effect of $\Gamma$ on the phase-diagram for BNP/RBNP
can be tied to the effect of varying the mixture volume fraction $\phi$.
This is shown in  Fig. \ref{fig5}
where we have plotted three curves for different 
volume fractions ($\phi=0.1, 0.3, 0.5$) with $\phi_l/\phi_s=10^{-8}$.
We observe that the range of size- and mass-ratios,
for which the  RBNP exists, increases with decreasing $\phi$.
This implies that the possibility of RBNP will increase
with increasing shaking strength  $\Gamma$.
The experimental findings of Breu {et al.} \cite{BEKR03} 
showed similar trends with $\Gamma$,
which our model is able to mimic.

To understand the effect of mean volume fraction on the BNP/RBNP
transition, we analyse different segregation forces 
for a given size-ratio and mass-ratio
(say, at $r_l/r_s=4$ and  $m_l/m_s=40$  in Fig. \ref{fig5}).
For this, both the Archimedean buoyancy force $F_b^A$
and the static geometric force $F_{ge}^{st}$ 
do not vary with $\phi$.
The dynamic geometric force $F_{ge}^{dyn}$, however, decreases with
decreasing $\phi$. In this limit, the collision frequency decreases
that decreases the collisional pressure,
and hence the intruder will become relatively less mobile. 
For this case, even though the granular energy ratio do not vary appreciably
with $\phi$, the pseudo-thermal
buoyancy force $F_b^T$ will decrease with $\phi$
as a result of a decrease in the value of ${Z_l}/{Z_s}$.
Both these effects result in 
a lower value of the `upward' force on the intruder
and hence the intruder will show the RBNP
with decreasing mean volume fraction (i.e. with increasing
shaking strength).
For these parameter conditions, the mean density at which the
BNP/RBNP transition occurs is about $0.284$.

\subsection{{\label{EIVF}}Effect of Intruder Volume Fraction}

So far we have shown results for a binary mixture in the tracer
limit of intruders (i.e. $\phi_l/\phi_s =10^{-8}$).
Recall from Fig. \ref{fig1} that the zone of RBNP shrinks to zero
as we make the particles more and more dissipative
since the pseudo-thermal buoyancy force increases in the same limit.
Hence, the RBNP is unlikely to occur in the tracer limit
for moderately dissipative particles.

\begin{figure}
\includegraphics[width=10.0cm]{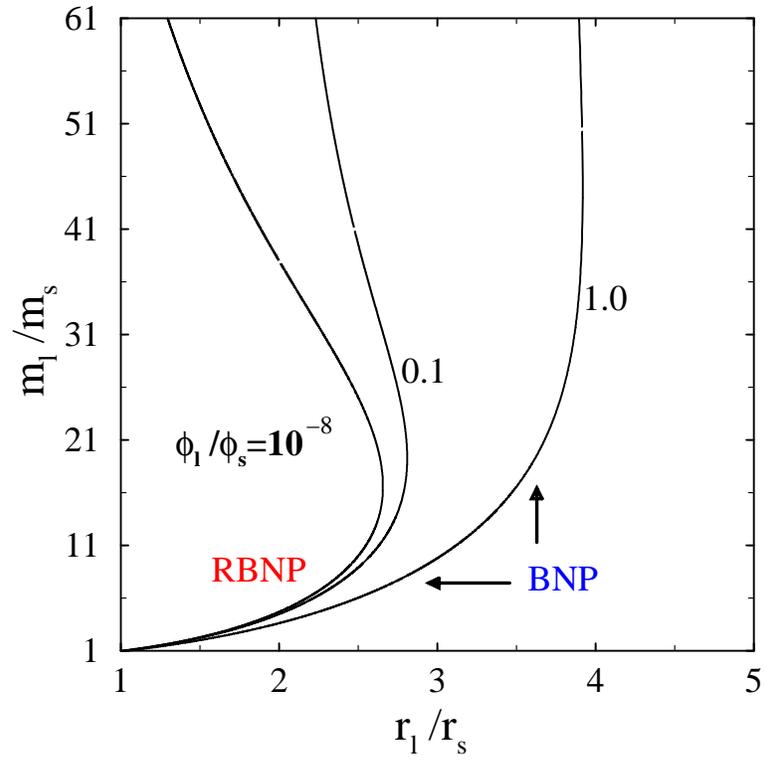}\\
\caption{
(color online)
Effect of the relative volume fraction of intruders on the 
phase diagram for BNP/RBNP in three dimensions:
$\phi=0.5$ and $e=0.95$.
}
\label{fig6}
\end{figure}

To probe the effect of the volume fraction of the intruders, 
we show the phase-diagram for the BNP/RBNP transition in Fig. \ref{fig6} for three
relative volume fractions: $\phi_l/\phi_s =10^{-8}$, $0.1$ and $1$.
The other parameters are as in Fig. \ref{fig2}.
As we increase the relative volume fraction of the intruders, the
ranges of mass- and size-ratios for which the reverse segregation (RBNP)
occurs increase sharply.
It is interesting to recall that 
most of the experiments of Breu {\it et al.} \cite{BEKR03} correspond
to the case where the number of layers of each species was equal;
this translates into $\phi_l > \phi_s$.
However, when they {\it reduced the number density of larger particles
they could not observe RBNP} (with other conditions remaining the same).
Also, the recent experiments of Yan {\it et al.} \cite{Yan03}
(at reduced air-pressures) do not show RBNP.
Our model predictions are, therefore, in qualitative agreement with these
experimental findings.
Clearly, more experiments are needed to map out the
correct phase-diagram for various volume fractions of the intruders.

\section{{\label{DSI}}Dynamics of Intruders in Tracer Limit: Nonmonotonic Rise Time}
                                                                                           
As an application of our theoretical framework,
we now consider the dynamics of intruders in the tracer limit ($n_l<< n_s$)
and determine the intruder's relative
velocity $u_l^r$ and its {\it rise-time}.
(Hereafter, the dynamics of intruder particles in the tracer limit
is considered to be that of a single intruder in a bed of small particles.)
Note that these two quantities
have been measured \cite{DU,CWHB96,HR04} in many Brazil-nut experiments.

To probe the motion of the intruder, we need to know
the macroscopic velocity-field $u_s$ of the smaller particles {\it a priori}.
Let the system be excited by a periodic force in the vertical
direction with a (symmetric) harmonic displacement:
   $z_p(t) = A\sin(\omega t)$.
We assume  that the vibrofluidized state of the smaller
particles is coupled to this periodic movement,
and make the following approximation for 
the velocity $u_s$ of the smaller particles:
\begin{equation}
   u_s\thickapprox \frac{\mathrm{d}z_p}{\mathrm{d}t} = - A\omega\cos(\omega t). 
\label{approx}
\end{equation}
This simply implies that 
the smaller particles follow the motion of the base plate
which is a reasonable  assumption  for shakings at {\it a high frequency
but with low amplitudes} ($\Gamma >>1$).

We  need to estimate the granular temperature of the mixture
$T$ which appears in the drag term in the evolution equation (\ref{dynamics}).
In the tracer limit  ($n_l<<n_s$), the mixture granular temperature
would be that of an equivalent mono-disperse system
of smaller particles \cite{SS82}; we have already assumed that the
granular temperature is uniform throughout the bed.
Considering this  homogeneous-state of  fluidization,
the global temperature can be estimated by equating the rate
of energy input through the bottom plate with
the rate of energy loss due to inelastic
particle collisions \cite{McNL98}:
\begin{equation}
   T = \frac{1}{\sqrt{2}\pi}\frac{m_s S_p}{N r_s^2}\frac{(A
       \omega)^2}{(1-e_{ss}^2)}.
\label{eqn_temp1}
\end{equation}
Here $S_p$ is the surface area of the base-plate and $N$ is the
total number of particles.
As demonstrated in many earlier studies,
this provides a reasonable approximation for the 
average temperature of the bed, and represents the leading-order solution 
even in the dense limit \cite{McNL98}.
We assume that in the fluidized state the height of the bed is $H$.
Thus, the expression for the global temperature
can be rewritten as
\begin{equation}
  T = \frac{\sqrt{2}}{3}\frac{m_s A^2\omega^2}{\phi_s R_H R_{ls}(1-e_{ss}^2)}.
\label{eqn_temp2}
\end{equation}
where
\begin{equation}
   R_H = \frac{H}{2r_l}
\label{par_3}
\end{equation}
is the non-dimensional bed-height as a multiple of the
intruder-diameter, and $R_{ls}=r_l/r_s$ the size-ratio.

Let us introduce the following reference scales for non-dimensionalization:
{\large
\begin{equation}
\left.
\begin{array}{rcl}
 \overline{u_l^r} &=& \frac{u_l^r}{u_R} = \frac{u_l^r}{A\omega} \\
 \overline{T} &=& \frac{T}{T_R} = \frac{T}{m_s A^2\omega^2}\\
 \overline{t} &=& \frac{t}{t_R} = \frac{t}{\omega^{-1}}
\end{array}
\right\}
\end{equation}
}
where the quantities with overbars are non-dimensional.
Hereafter, for convenience, we will drop the overbar on the non-dimensional quantities.
With this scaling, the non-dimensional granular temperature 
has the following expression:
\begin{equation}
  T = \frac{\sqrt{2}}{3\phi_s R_H R_{ls}(1-e_{ss}^2)}.
\label{eqn_temp_nd}
\end{equation}

The evolution equation (\ref{dynamics}) 
in non-dimensional form can now be written as
\begin{equation}
   \frac{\mathrm{d}u_l^r}{\mathrm{d}t}= \alpha\left(\Gamma^{-1} - \sin{t}\right)
             - \beta u_l^r
\label{eqn_evol_nd}
\end{equation}
where 
\begin{equation}
  \beta = \beta_0 R_A\sqrt{T}
\end{equation}
is the non-dimensional drag coefficient which 
is a function of the bed-temperature $T$ and the amplitude
of the harmonic-shaking
\begin{equation}
  R_A = \frac{A}{2 r_l} .
\label{par_2}
\end{equation}
Here $\alpha$ and $\beta_0$ are non-dimensional functions of 
the size-ratio, mass-ratio and volume fraction:
\begin{eqnarray*}
  \alpha &=& \left(\frac{m_s}{m_l}\frac{Z_l}{Z_s}\frac{T_l}{T_s} - 1\right) \\ 
  \beta_0 &=& 8K_{ls}\left(\frac{R_{ls}}{R_{ls} + 1}\right)
       \left(\frac{2}{\pi}\frac{m_s}{m_l}M_{sl}\right)^{1/2} \\
   K_{ls} &=& \frac{1}{4}\phi_s\chi_{ls}(1+R_{ls})^3 \\
   M_{sl} & =& \frac{m_s}{m_l+m_s} \\
   R_{ls} &=& \frac{r_l}{r_s} .
\end{eqnarray*}

The solution to the differential equation (\ref{eqn_evol_nd}), 
with initial condition
$u_l^r(t)|_{t=0}=0$, is given by
\begin{equation}
  u_l^r(t) = \frac{\alpha}{\beta\Gamma}\left(1 - e^{-\beta t}\right) 
    -    \left(\frac{\alpha}{1+\beta^2}\right)
          \left(e^{-\beta t} - \cos{t} +\beta\sin{t}\right) .
\label{eqn_vel1}
\end{equation}
The temporal evolution of 
the position of the intruder, $z_l(t)$, can be
obtained by integrating Eq. (\ref{eqn_vel1}):
\begin{eqnarray}
  z_l(t) &=& \frac{\alpha R_A}{\beta^2\Gamma}\left(\beta t + e^{-\beta t}\right) 
     - \frac{\alpha R_A}{\beta^2\Gamma}\left(1 + \beta\Gamma\right) + z_0 \nonumber \\
    &+ &  \left(\frac{\alpha R_A}{\beta(1+\beta^2)}\right)
          \left(e^{-\beta t} + \beta\sin{t} +\beta^2\cos{t}\right)
\label{eqn_pos1}
\end{eqnarray}
with $z_0\equiv z_l(0)$ being the position of the intruder at time $t=0$.
Note that $z_l$ has been non-dimensionalized by the diameter of the intruder.

\subsection{{\label{NRT}}Rise Time: Non-monotonicity and Experiments}

By knowing the steady relative  velocity of the intruder
$u_l^r$, we can calculate its
{\it rise-time} $\tau$:
\begin{equation}
  \tau = \frac{2 r_l}{u_l^r} = \frac{2 r_l\beta\Gamma}{\alpha} .
\label{eqn_rtime}
\end{equation}
This is the (non-dimensional) time that the intruder will take
to travel a height of one intruder diameter. 
In the following we have plotted the rise-time in cycles.

\begin{figure}
\includegraphics[width=10.0cm]{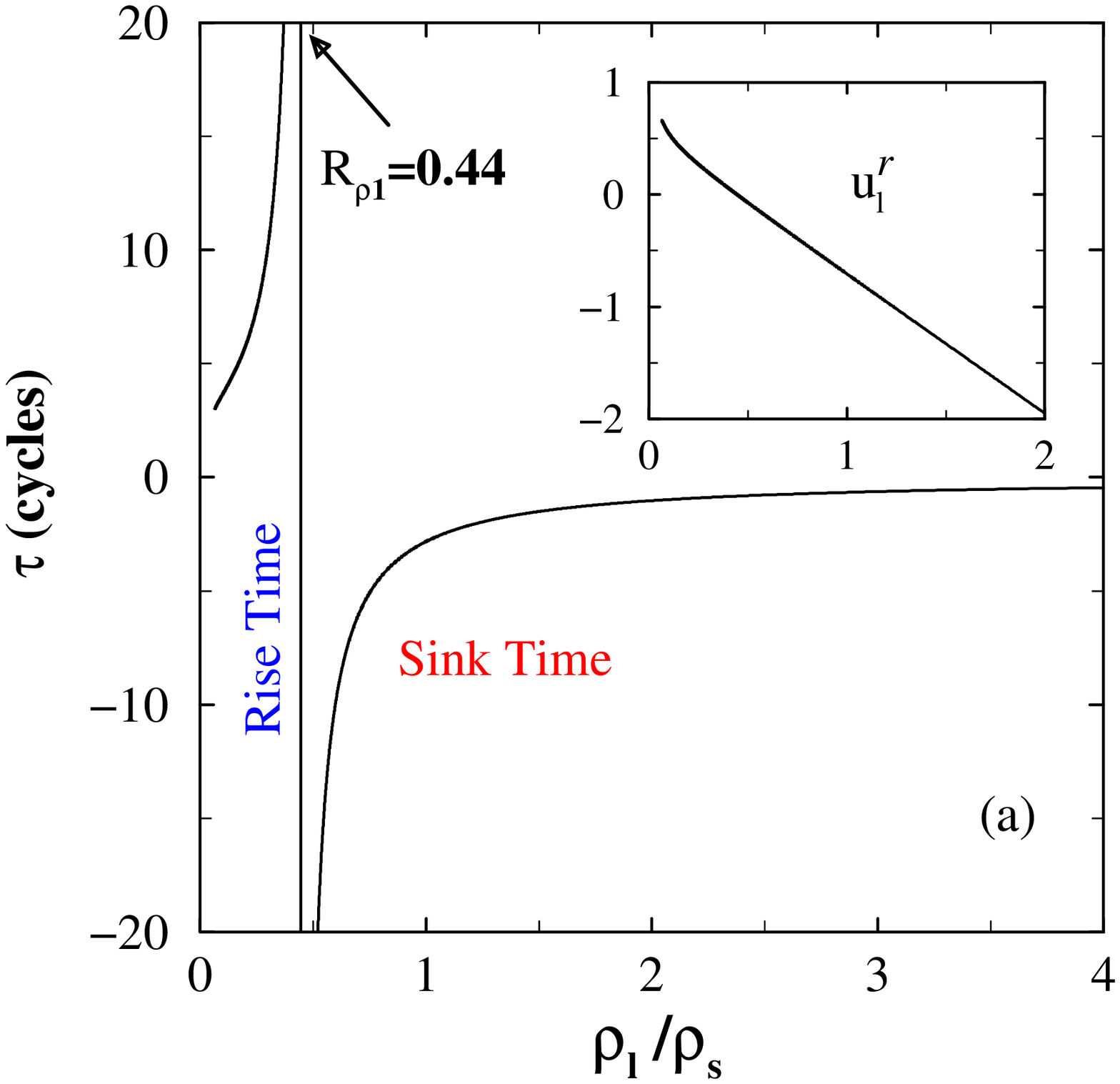}\\
\includegraphics[width=10.0cm]{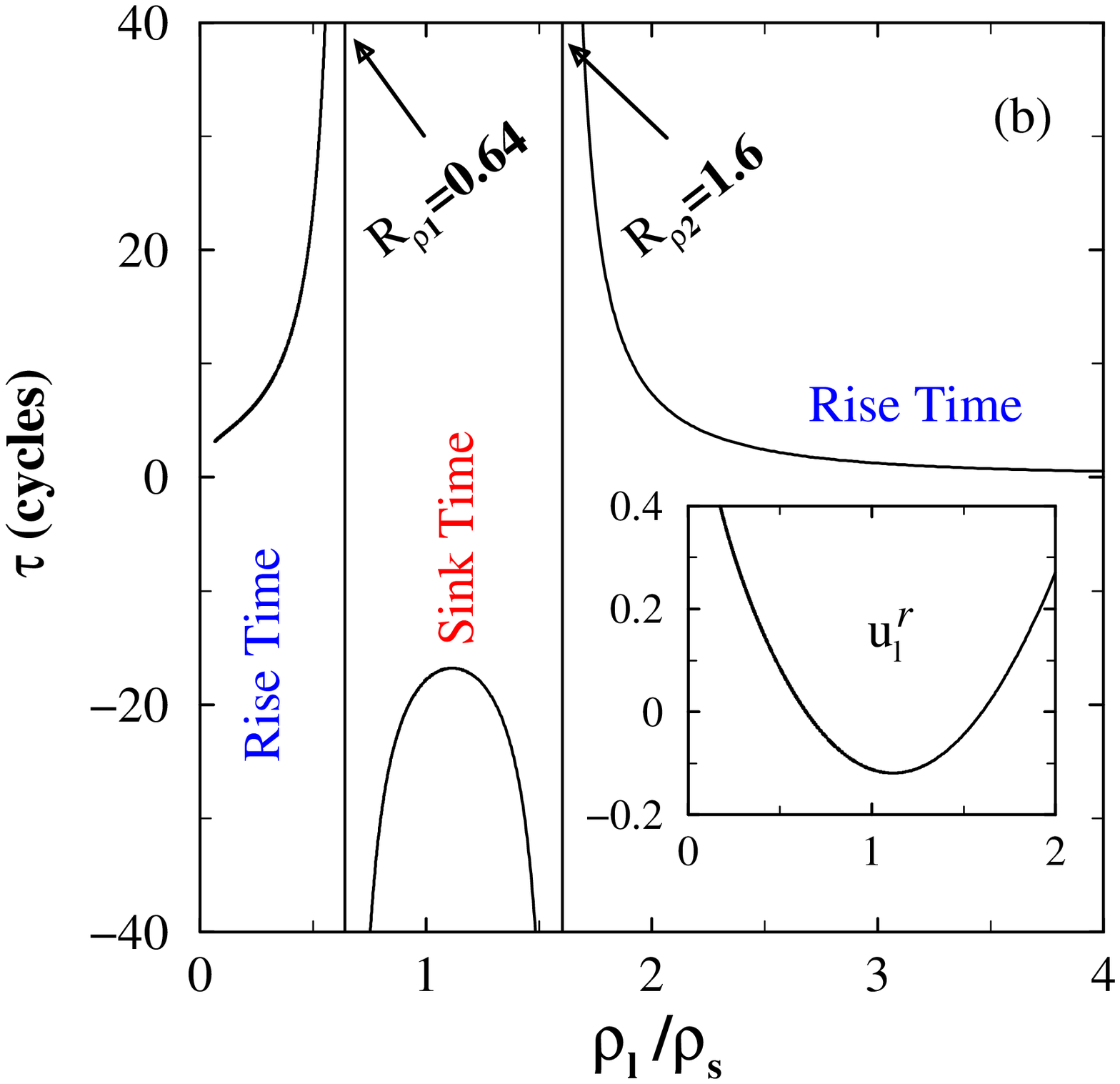}\\
\caption{
(color online)
(a) Variations of rise (/sink) time $\tau$ (in cycles)
with density-ratio for $T_l=T_s$; the parameter values
are as in Fig. \ref{fig4}. The shaking parameters are
$\Gamma=5$, $f=50$ Hz; the intruder diameter is $4.97$cm and
the bed-height $R_H=H/(2r_l)=10$.
(b) Same as $a$ with non-equipartition assumption.
The inset in each panel shows the variation of 
the intruder's relative velocity, $u_l^r$, with the density-ratio.
The vertical lines in each panel refer to 
density-ratios for the $BNP/RBNP$ transition.
}
\label{fig7}
\end{figure}

First, we consider an intruder of a given size-ratio
($R_{ls}=r_l/r_s=2.5$) and vary its density;
further, we assume the equipartition assumption $T_l=T_s$.
For this case, the variation of the rise-time with the density-ratio,
$R_\rho = \rho_l/\rho_s$, is shown in Fig. \ref{fig7}$a$;
other parameters are as in Fig. \ref{fig4}. 
The parameters for base-plate excitations are
$\Gamma=5$ and $f=50$ Hz; we set 
the bed-height to $R_H=H/(2r_l)=10$.
As expected, $\tau$ diverges at the onset of BNP/RBNP ($R_{\rho1}\approx 0.64$)
since the intruder velocity, $u_l^r$ changes its sign at this
threshold density as seen in the inset of Fig. \ref{fig7}$a$.
For $R_\rho> R_{\rho1}$, the sink-time decreases, implying that the heavier intruders
sink faster; for $R_\rho< R_{\rho1}$,
the lighter intruders rise faster.

Now we relax the equipartition assumption, i.e. $T_l\neq T_s$,
and plot the variation of $\tau$ with $R_\rho$ in Fig. \ref{fig7}$b$.
In contrast to the case in Fig. \ref{fig7}$a$,
now we have the pseudo-thermal buoyancy force
acting on the intruder. This additional upward force makes the
intruder to rise again beyond $ R_\rho> R_{\rho2}$.
Interestingly, the rise-time for $R_\rho> R_{\rho2}$ decreases
with increasing density-ratio, implying that
{\it the heavier the intruder the faster it rises}.
Such an effect has been observed in  experiments \cite{EXPNM,Yan03,HR04}.
Within the window of the RBNP ($R_{\rho1} < R_\rho> R_{\rho2}$),
we observe that the sink-time 
varies non-monotonically and diverges at both the limits.
For this parameter combination, an intruder with $R_\rho=1$ will
take about $15$ cycles of excitation to travel (sink)
a distance of its diameter.

\begin{figure}
\includegraphics[width=10.0cm]{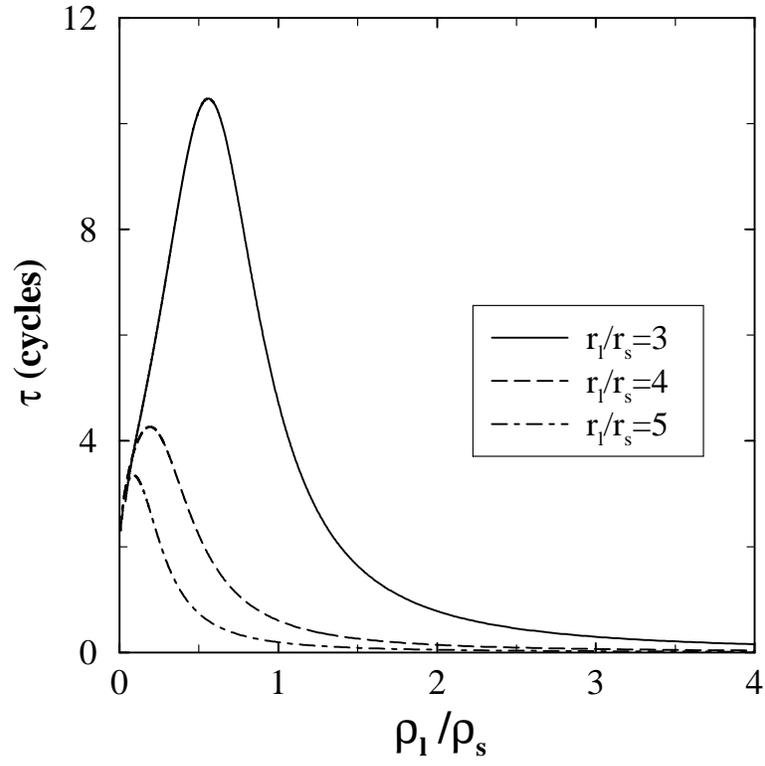}
\caption{
Variation of rise time $\tau$ (in cycles)
with density-ratio for different size-ratios; the parameter values
are as in Fig. \ref{fig5}. The shaking parameters are
as in Fig. \ref{fig7}.
}
\label{fig8}
\end{figure}

With other parameter conditions as in Fig. \ref{fig7}$b$,
we increase the intruder size such that we are in the zone of BNP
(see Fig. \ref{fig1}). This is shown in  Fig. \ref{fig8}
for three different size-ratios ($R_{ls}=3, 4, 5$).
It is interesting to note that the rise-time
is {\it non-monotoic} with the density-ratio. For
a given size-ratio, {\it there is a threshold density-ratio ($R_\rho<1$)
for the intruder above/below which it rises faster}.
This threshold density-ratio shifts to a lower value
with increasing size-ratio. 
Note further that at a given density-ratio {\it the larger the intruder
the faster it rises} in conformity with our earlier observation
in Fig. \ref{fig7}$b$.

\begin{figure}
\includegraphics[width=10.0cm]{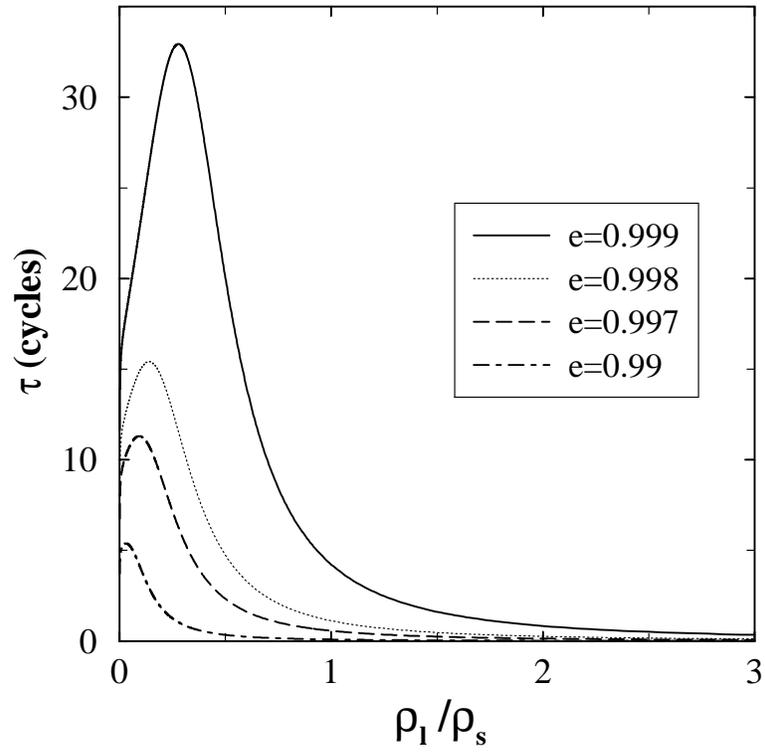}
\caption{
Variation of rise time $\tau$ (in cycles)
with density-ratio for different restitution coefficients.
The size-ratio is set to $r_l/r_s=10$;
other parameters as in Fig. \ref{fig8}.
}
\label{fig9}
\end{figure}

Next we fix a large size-ratio of $R_{ls}=10$,
and show the variation of $\tau$ with the density ratio 
$R_\rho$ in Fig. \ref{fig9} for four different restitution coefficients
($e=0.999, 0.998, 0.997, 0.99$).
With increasing dissipation, the magnitude of the
energy-ratio, $T_l/T_s$, increases that generates
additional pseudo-thermal buoyancy force on
the intruder, thereby increasing its velocity.
Hence the intruder will move faster and
the height of the peak in the rise-rate curve 
decreases as observed in Fig. \ref{fig9}.

\subsection{{\label{RBNP}}Comparison with Experiments}

In order to verify the possible effects of convection \cite{KJN93},
here we make a qualitative comparison of 
our predictions on the rise time with
the recent experiment results of Huerta and Ruiz-Suarez \cite{HR04}.
They performed two-sets of experiments to test the effects of bulk convection,
one at $f=5 Hz$ and $\Gamma=3$ (i.e. the low frequency but
high amplitude limit), and the other at $f=50 Hz$ and $\Gamma=3$
(i.e. the high frequency but low amplitude limit).
(They verified that  
the  bulk-convection was negligible in the high frequency limit, 
but  was present in the  low frequency limit.)
They observed that the rise-time is {\it non-monotonic} in 
the  high-amplitude (and low frequency) limit (see their Fig. 1),
but in the absence of convection the heavier particles sank
(see their Fig. 3).

Note that most of the assumptions
in our model are in tune with their  high frequency experiments
since the convection effects are claimed to be  negligible in that limit.
However, if the  level of pseudo-thermal buoyancy force is small,
the window of the RBNP ($R_{\rho1} < R_\rho < R_{\rho2}$, refer to
Fig. \ref{fig7}$b$) can be made arbitrarily large as in Fig. \ref{fig7}$a$.
In this case, the Archimedean buoyancy force competes
with geometric forces, thereby  deciding the dynamics
of the intruder for $R_\rho>R_{\rho1}$ as in our Fig. \ref{fig7}$a$
which looks remarkably similar to Fig. 3 in ref. \cite{HR04}.

Moving onto their low frequency experiments, we need
to consider the additional effects of bulk-convection.
Note that the bulk-convection generates an upward-plume at the 
center of the container, and two `narrow' downward plumes near the side-walls;
clearly, the  granular energy of the particles within the center-plume 
will be much greater than that of bulk material, and this
would lead to another competing force on the intruder.
Under the present formalism,  this additional force would 
{\it reinforce} the pseudo-thermal buoyancy force.
Thus, the {\it effective}  pseudo-thermal buoyancy force
will increase in the presence of convection, leading to an
over-turning of the phase-coexistence line 
beyond a critical size-ratio (refer to  our Fig. \ref{fig1}).
Hence the added effect of bulk-convection
is to take us to the regime of BNP and the heavier particles will also rise.
This is precisely what we see in Fig. 1 of ref. \cite{HR04}.
Therefore, in their low frequency experiments,
the {\it effective}  pseudo-thermal buoyancy force
would compete with the Archimedean buoyancy force and geometric forces.

Lastly, we comment on the experiments of M\"obius {\it et al.} \cite{EXPNM}
who first showed  that the rise-time of an intruder ($r_l/r_s >10$)
in a bed of small particles ($r_s=250\mu m$) 
is non-monotonic with the density-ratio.
They  also showed that the peak on the rise-time curve decreases in height and
shifts to a lower density with decreasing the  air-pressure,
and vanishes as the air-pressure approaches $1$ torr.
It is conceivable that the decreased air-pressure has reduced the {\it effective}
drag on the intruder and thereby increased its velocity--
hence the height of the peak (maximum rise-time) decreases
with decreasing air-pressure.
But the shifting of this  peak to a lower density
remains unexplained.

\section{{\label{DIS}}Discussion}

\subsection{{\label{sRBNP}}A Simple Explanation for RBNP: Boltzmann Limit}

The reverse segregation  effect can be understood by considering the Boltzmann-limit 
(i.e. the dilute limit $\phi\to 0$) for
which the equation of state of species $i$ is
\[
   p_i = n_i T_i.
\]
By integrating the steady vertical momentum equation, we obtain
an expression for the number density profile of species $i$, 
\begin{equation}
   n_i(z) 
          = n_i(0)\exp\left[-\frac{m_igz}{T_i}\right],
\end{equation}
where we have assumed that $T_i$ is independent of $z$.
(This assumption implies that the variation of number-density with $z$
is due to the variation of partial pressure along $z$.)
Hence,  the ratio of number densities is
\begin{equation}
   \frac{n_l(z)}{n_s(z)} = \frac{n_l(0)}{n_s(0)}\exp\left[-\frac{m_l}{T_l}
      \left(1 - \frac{T_l}{T_s}\frac{m_s}{m_l}\right)gz\right] .
\end{equation}

For a mixture of particles with purely elastic collisions, 
the equipartition principle holds, i.e., $T_l=T_s$.
In this case, the particles segregate according to
their masses, driven by the Archimedean buoyancy force,
and there is {\it no} segregation if the particles are of the {\it same} mass.
Thus, the heavier particles remain at the bottom and lighter particles
at the top, leading to the RBN effect.

Now considering {\it inelastic} particles 
of equal masses $m_l=m_s$ but having different sizes and densities,
we know that $T_l<T_s$ for all values of restitution coefficient.
This implies that the decay-rate of $n_l$ with $z$ is {\it faster} than that of $n_s$.
Hence the center of mass of the larger-particles 
will be at a lower-level than
that of the smaller-particles, 
\[
   \frac{<z_l>_{cm}}{<z_s>_{cm}} < 1,
\]
clearly leading to the RBN effect.

\subsection{{\label{HOE}}Higher-order Effects and Possible Experiments}
  
The granular hydrodynamic model is a natural and convenient
framework for studying fluidized systems.  It is nonetheless
criticized for being overly simplified and  unrealistic \cite{JS83}.
In our analysis  we have considered the hypothesis of a Maxwellian
velocity distribution function. Let us remark that the
leading--order solution of the Boltzmann--Enskog kinetic equation
for granular mixtures is a Maxwellian (see ref. \cite{JM87}), and
the non--Gaussian correction term remains relatively small \cite{PCMP03}.
More importantly, since the present theory is restricted
to the Euler-level description of hydrodynamic equations,  
we do not need to take into account the corrections
due to non-Maxwellian effects as explained in \cite{TAH03}. 
Moreover, the Archimedean buoyancy force ($F_{b}^{A}$) and the static geometric
force ($F_{ge}^{st}$) do not depend on the distribution function,
and the non--Gaussian correction does not affect the dynamic
geometric force ($F_{ge}^{dyn}$). Only the pseudo-thermal buoyancy force
($F_{b}^{T}$) can be expressed in terms of the velocity
distribution function. Thus, the higher-order corrections would
affect this buoyancy force-- a detailed investigation
of this is left for a future investigation.

All the segregation forces are indirectly measurable in a standard
vibrated--bed setup, and thereby making them  directly verifiable via
experiments. For example, the thermal buoyancy force ($F_b^T$) can
be measured by measuring the granular energies ($T_l$ and
$T_s$) from the snapshots of successive particle configurations.
These snapshots can also be used to measure the pair correlation
function ($\chi_{ij}$ and hence the compressibility factor $Z_i$) and
the dynamic geometric force ($F_{ge}^{dyn}$). The other two forces
($F_b^{A}$ and $F_{ge}^{st}$) are then easily obtained. 
Knowing all these forces from experiments, one could construct
each term of our theory independently.
We hope that the present investigation will stimulate 
future experiments to measure these quantities.

\section{{\label{CON}}Summary and conclusions}
                                                                                           
One of our motivations has been to explore whether the
hydrodynamic--description of granular materials can be applied to
the derivation of a time--evolution equation for the segregation
velocity of intruders in a dense fluidized bed. 
We have done that in our previous paper \cite{TAH03},
starting from the kinetic-theory of binary mixtures. 
Based on this time--evolution equation, 
here we have tried to provide a unified  theoretical
description for the segregation dynamics of Brazil-nuts in a vibrated granular
mixture.

We have discussed a novel mechanism for segregation and 
argued that the onset of segregation is due to a
competition between the {\it buoyancy} and {\it geometric} forces.
Apart from the standard Archimedean buoyancy force, we have 
introduced the notion of a pseudo-thermal buoyancy force
that results from the fact that the intruder particle
fluctuates at a {\it different} energy level than
the smaller particles. (The microscopic dissipation, 
due to the inelastic nature of particle-collisions, is responsible for
the non-equipartition of granular energy.)
The size-difference between the intruder and the bed-material
results in two geometric forces, 
one compressive and the other-one tensile in nature-- the net geometric force
is always compressive.
Thus, the net buoyancy force competes with the net compressive
geometric force, thereby deciding the onset of BNP/RBNP transition.

For a mixture of perfectly hard-particles with elastic collisions,
the pseudo-thermal buoyancy force is zero but the intruder has to
overcome the net compressive geometric force to rise.
For this case, the competition between 
the net geometric force and  the  Archimedean buoyancy force
yields a threshold density-ratio, $R_{\rho 1}=\rho_l/\rho_s < 1$,
above which the {\it lighter intruder sinks},
thereby signalling the {\it onset} of the {\it reverse buoyancy} effect.
For a mixture of dissipative particles, on the other hand,
the non-zero pseudo-thermal buoyancy force
gives rise to another threshold density-ratio, $R_{\rho 2}$ ( $> R_{\rho 1}$),
above which the intruder rises again.

This theory is applied to study the dynamics of
a single intruder (i.e. in the tracer limit of intruders) 
in a vibrofluidized granular mixture,
with the effect of the base-plate excitation
being taken into account through a `mean-field' assumption.
The rise (/sink) time {\it diverges} near the threshold density-ratio
for reverse-segregation. The most interesting result is 
that the rise-time of the intruder could vary {\it non-monotonically}
with the density-ratio. 
For a given size-ratio (in the zone of BNP), there is a threshold density-ratio
for the intruder at which it takes maximum time to rise and 
above(/below) which it rises faster.
This implies that {\it the heavier (and larger) the intruder, the faster it ascends};
a similar effect has been observed in some recent experiments \cite{EXPNM,HR04}.
The peak on the rise-time curve decreases in height and shifts
to a lower density-ratio as we increase the 
magnitude of the pseudo-thermal buoyancy force.

The  main message emerging from our
analysis is that the pseudo-thermal buoyancy force that
results from the non-equipartition of granular energy ($T_l\neq T_s$) 
plays an important role in the segregation process.
Even though the source of this energy non-equipartition is inelastic dissipation,
it can be argued  that the effects of the bulk-convection
will also lead to a 
separation between the granular energies of the intruder
and the bed-particles even in the limit of perfectly elastic collisions. 
Thus, the increased pseudo-thermal buoyancy force
(i.e. with increasing dissipation-levels in our model)
can also be related to the presence of the bulk-convective motion. 
In this connection it may be pointed out that a different functional form 
for the equation of state might change the precise threshold of
segregation, but the presence of additional pseudo-thermal buoyancy force,
for example, due to convection, is likely to retain
the overall phase-diagram similar.

It is remarkable that our theory can explain  several
experimental observations in a unified manner,
despite having many simplifying assumptions. 
In addition to explaining the experimental
results of Breu {\it et al.} \cite{BEKR03} on 
the reverse Brazil-nut segregation,
we have suggested a plausible explanation 
for the {\it origin} of the onset of the reverse-buoyancy
effect of Shinbrot and Muzzio \cite{SM98} as well as
the {\it origin} of the non-monotonic ascension-dynamics \cite{EXPNM,HR04}.
However, the effects of air-pressure on the non-monotonic ascension-dynamics
\cite{EXPNM}  seems to be  more  subtle and needs further work.
This can be done by considering a three-phase mixture, with separate
balance equations for the air and taking into account the resulting
interactions on both the smaller particles and the intruders.
Lastly, the effects of Coulomb friction 
and the particle roughness need to be incorporated in our model in future.

\begin{acknowledgments}
The work on `reverse segregation' was started when M.A. was
a Humboldt fellow at ICA1, Stuttgart, and L.T. was a doctoral
student at PMMH, ESPCI, Paris.
The idea on `reverse buoyancy' and `non-monotonic ascension dynamics' was
pursued at JNCASR.
M.A. acknowledges computational facilities and the financial support from 
the JNCASR (Grant Number: PC/EMU/MA/35)
as well as the hospitality of ICA1 (funded by AvH Fellowship).
\end{acknowledgments}



\newpage

\newpage

\newpage

\newpage

\newpage

\newpage
\vspace*{2.0cm}

\newpage

\newpage
\vspace*{2.0cm}

\newpage
\vspace*{2.0cm}

\end{document}